\documentclass[smus]{snow2e}
\usepackage{graphicx}

\catcode`@=11
\def\tableofcontents{\@starttoc{toc}}

\renewcommand*\l@section[2]{%
  \ifnum \c@tocdepth >\z@
    \addpenalty\@secpenalty
    \addvspace{1.0em \@plus\p@}%
    \setlength\@tempdima{2.5em}
    \begingroup
      \parindent \z@ \rightskip \@pnumwidth
      \parfillskip -\@pnumwidth
      \leavevmode \bfseries
      \advance\leftskip\@tempdima
      \hskip -\leftskip
      #1\nobreak\hfil \nobreak\hb@xt@\@pnumwidth{\hss #2}\par
    \endgroup
  \fi}

\renewcommand*\l@subsection{\@dottedtocline{2}{2.5em}{2.0em}}
\renewcommand*\l@subsubsection{\@dottedtocline{3}{4.5em}{2.0em}}

\catcode`@=12

\def\ar#1#2#3{     {\it Ann. Rev. Nucl. and Part. Sci. }{\bf #1}, #2 (#3)}

\def\np#1#2#3{           {\it Nucl. Phys. }{\bf #1}, #2 (#3)}
\def\pl#1#2#3{           {\it Phys. Lett. }{\bf #1}, #2 (#3)}
\def\pr#1#2#3{           {\it Phys. Rev. }{\bf #1}, #2 (#3)}

\def\prl#1#2#3{          {\it Phys. Rev. Lett. }{\bf #1}, #2 (#3)}   

\def\rmp#1#2#3{          {\it Rev. Mod. Phys. }{\bf #1}, #2 (#3)}

\def\etal{\hbox{\it et al.}}
\def \pb {\rm pb}

\def\etal{\hbox{\it et al.}}
\def \GeV {\rm GeV}
\def \TeV {\rm TeV}
\def \to {\rightarrow}

\def\dofig#1{\centerline{\includegraphics[width=8.5cm]{#1}}}

\def\dofigb#1{\centerline{\includegraphics[width=2.5cm]{#1}}}
\def\dofigc#1{\centerline{\includegraphics[width=12cm]{#1}}}

\def\dofige#1{\centerline{\includegraphics[width=10cm]{#1}}}
\def\dofigf#1{\centerline{\includegraphics[width=7cm]{#1}}}

\def \et {E_{T}}

\def\cmsec{{\rm cm}^{-2}s^{-1}}

\def\simge
{\mathrel{\rlap{\raise 0.53ex \hbox{$>$}}{\lower 0.53ex \hbox{$\sim$}}}}
\def\simle
{\mathrel{\rlap{\raise 0.4ex \hbox{$<$}}{\lower 0.72ex \hbox{$\sim$}}}}

\def\slashchar#1{\setbox0=\hbox{$#1$}           
   \dimen0=\wd0                                 
   \setbox1=\hbox{/} \dimen1=\wd1               
   \ifdim\dimen0>\dimen1                        
      \rlap{\hbox to \dimen0{\hfil/\hfil}}      
      #1                                        
   \else                                        
      \rlap{\hbox to \dimen1{\hfil$#1$\hfil}}   
      /                                         
   \fi}                                         %

\def\Etmiss{\slashchar{E}_T}

\catcode`@=11
\def\citenum#1{%
   \expandafter\ifx\csname b@#1\endcsname\relax{\bf ??}\else
   \csname b@#1\endcsname\fi}
\catcode`@=12

\def \gluino {\tilde{g}} 
\def \squark {\tilde{q}}

\def\s{\tilde}

\def\tq{{\tilde q}}
\def\tchi{{\tilde\chi}}

\def\lsp{{\tilde\chi_1^0}}
\def\tG{{\tilde G}}
\def\ttau{{\tilde\tau}}
\def\tell{{\tilde\ell}}
\def\neu#1{\tilde\chi^0_{#1}}

\def\GeV{{\rm GeV}}
\def\TeV{{\rm TeV}}
\def\Meff{M_{\rm eff}}

\def\pb{\rm pb}
\def\fb{\rm fb}

\def  \met {\not\!\!\et }

\def  \abseta {|\eta|}

\def \fbi{{\rm fb^{-1}}}

\def\jet{{\rm jet}}
\def\jets{{\rm jets}}

\begin{document}

\title{ \hfill  BNL-HET-01/33\\
High Transverse Momentum Physics at the Large Hadron
Collider}
\author{{\bf The ATLAS and CMS Collaborations}\\
Edited by\\
J.G. Branson$^{\rm a}$, D. Denegri$^{\rm b}$, I. Hinchliffe$^{\rm c}$, 
F. Gianotti$^{\rm d}$, F.E. Paige$^{\rm e}$, and P. Sphicas$^{\rm d,f}$\\
$^{\it a}${\it U. of California San Diego, La Jolla, CA 92093}\\
$^{\it b}${\it DAPNIA, CEA/Saclay, F-91191 Gif-sur-Yvette, France}\\
$^{\it c}${\it Lawrence Berkeley National Laboratory, Berkeley, CA 94720}\\
$^{\it d}${\it CERN, CH-1211 Geneva 23, Switzerland}\\
$^{\it e}${\it Brookhaven National Laboratory, Upton, NY 11973}\\
$^{\it f}${\it MIT, Cambridge, MA 02139}\\
}
\setlength{\titleblockheight}{7.55cm}
\maketitle
\tableofcontents
\pagestyle{plain}

\begin{abstract}

This note summarizes many detailed physics studies done by the ATLAS and CMS
Collaborations for the LHC, concentrating on processes involving the
production of high mass states. These studies show that the LHC should be able
to elucidate the mechanism of electroweak symmetry breaking and to study a
variety of other topics related to physics at the TeV scale. In
particular, a Higgs boson with couplings given by the Standard Model
is observable in several channels over the full range of allowed
masses. Its mass and some of its couplings will be determined. If
supersymmetry is relevant to electroweak interactions, it will be
discovered and the properties of many supersymmetric particles
elucidated. Other new physics, such as the existence of massive gauge
bosons and extra dimensions can be searched for extending existing
limits by an order of magnitude or more\footnote{Note that this document lacks the detector
  pictures due to file size limiations of the Los Alamos archive. The
  version with pictures can be found at http://www-theory.lbl.gov/$\sim$ ianh/lhc}.

\end{abstract}
\section{Introduction and Motivation}

This document summarizes the potential of the Large Hadron Collider
(LHC) for high mass and high transverse momentum physics and explains
why the LHC is expected to provide a crucial next step in our
understanding of nature. The results given here are based on 
publically available work done
by many ATLAS and CMS collaborators either as part of the design of the
ATLAS~\cite{atlas} and CMS~\cite{cms} detectors or in subsequent
investigations. On the basis of these studies, we believe that the
physics potential of the LHC is enormous: among currently approved
projects in high energy physics, it uniquely has sufficient energy and
luminosity to probe in detail the TeV energy scale relevant to
electroweak symmetry breaking.

\subsection{The Standard Model}

The Standard Model (SM) is a very successful description of the interactions
of the components of matter at the smallest scales ($\simle 10^{-18}\,$m) and
highest energies ($\sim 200\,$GeV) accessible to current experiments.  It is a
quantum field theory that  describes the interaction of spin-$1\over 2$,
point-like fermions, whose interactions are mediated by spin-1 gauge bosons.
The existence of the gauge bosons and the form of their interactions are
dictated by local gauge invariance, a manifestation of the symmetry group of
the theory, which for the SM is $SU(3) \times SU(2) \times U(1)$.

The fundamental fermions are leptons and quarks; the left-handed states are
doublets under the $SU(2)$ group, while the right-handed states are singlets.
There are three generations of fermions, each generation identical except for
mass. The origin of this structure, and the breaking of generational symmetry
(flavor symmetry) remain a mystery.  There are three leptons with electric
charge $-1$, the electron ($e$), muon ($\mu$) and tau lepton ($\tau$); and
three electrically neutral leptons, the neutrinos $\nu_e$, $\nu_\mu$ and
$\nu_\tau$. Similarly there are three quarks with electric charge
$+{2\over3}$, up ($u$), charm ($c$) and top ($t$); and three with electric
charge $-{1\over3}$, down ($d$), strange ($s$) and bottom ($b$). The quarks
are triplets under the $SU(3)$ group and thus carry an additional ``charge,''
referred to as color. There is mixing between the three generations of quarks,
which in the SM is parameterized by the Cabibbo-Kobayashi-Maskawa
(CKM)~\cite{ckm} matrix but is not explained.

In the SM the $SU(2)\times U(1)$ symmetry group describes the electroweak
interactions. This symmetry is spontaneously broken by the existence of a
(postulated) Higgs field with a non-zero expectation value, leading to massive
vector bosons --- the $W^\pm$ and $Z$ --- which mediate the weak interaction;
the photon of electromagnetism remains massless. One physical degree of
freedom remains in the Higgs sector, a neutral scalar boson $H^0$, which is
presently unobserved. The $SU(3)$ group describes the strong interaction
(quantum chromodynamics or QCD). The eight vector gluons that mediate this
interaction themselves carry color charges and so are self-interacting. This
implies that the QCD coupling $\alpha_S$ is small for large momentum transfers
but large for small momentum transfers, and leads to the confinement of quarks
inside color-neutral hadrons. Attempting to free a quark produces a jet of
hadrons through quark-antiquark pair production and gluon bremsstrahlung. The
smallness of the strong coupling at large mass scales makes it possible to
calculate reliably cross sections for the production of massive particles at
the LHC.

The basic elements of the Standard Model were proposed in the 1960's and
1970's~\cite{standard-model}. Increasing experimental evidence of the
correctness of the model accumulated through 1970's and 1980's:
\begin{itemize}
\item observation of (approximate) scaling in deep inelastic scattering
experiments, showing the existence of point-like scattering centers inside
nucleons, later identified with quarks~\cite{dis};
\item observation of the $c$ and $b$ quarks~\cite{bandc};
\item observation of neutral weak currents from $Z$
exchange~\cite{neutral-currents};
\item observation of jet structure and three-jet final states resulting from
gluon radiation in $e^+ e^-$ and hadron-hadron collisions~\cite{3jets};
\item direct observation of the $W$ and $Z$ at the CERN $Sp\overline{p}S$ 
collider~\cite{wz}.
\end{itemize} 
Following these discoveries, ever more precise experiments at LEP and
SLC have provided verification of the couplings of quarks and leptons to
the gauge bosons at the level of 1-loop radiative corrections ($\sim
{\cal O}(10^{-3})$). Also, the top quark has been discovered at Fermilab
with a very large mass ($\sim 175\,$GeV)~\cite{topdisc}.

With the recent direct observation of the $\nu_\tau$~\cite{nutau}, only one
particle from the Standard Model has yet to be observed, the Higgs
boson. The Higgs is very important because it holds the key to the
generation of $W$, $Z$, quark and lepton masses.  

Some of the SM parameters, specifically those of the CKM matrix, are not
well determined. In particular, while  CP violation is accommodated
in
the SM 
through a phase in the CKM quark mixing matrix, it remains poorly
understood.  CP violation was first observed in K
decays~\cite{FitchCronin}. Recently, direct CP violation has been seen
in K decays~\cite{ktev}, and evidence for CP violation in $B \to \psi
K_s$ has been seen in $B$-factories~\cite{bfactory,sinbeta} and in
CDF~\cite{cdf-b}. More precise measurements over the next few years
should determine these parameters or demonstrate the SM cannot
adequately explain CP violation.

The minimal SM can only accommodate  massless neutrinos and hence no neutrino oscillations.
There is evidence for such oscillations from measurements  by
SuperKamiokande
of neutrinos
produced in the atmosphere and from a deficit in the flux
of electron neutrinos from the sun\cite{numass}. While it is easy to extend
the SM to include neutrino masses, understanding their small values seems to
require qualitatively new physics.

\subsection{Beyond the Standard Model}

The success of the Standard Model~\cite{standard-model}
 of strong (QCD), weak and electromagnetic
interactions has drawn increased attention to its limitations. In its simplest
version, the model has 19 parameters~\cite{cahn96}, the three coupling constants of 
the gauge theory $SU(3)\times SU(2)\times U(1)$, three lepton and six quark
masses, the mass of the $Z$ boson which sets the scale of weak interactions,
the four parameters which describe the rotation from the weak to the mass
eigenstates of the charge $-1/3$ quarks (CKM matrix).
All of these parameters are determined with varying errors. 
One of the two remaining parameters is the coefficient $\theta$ of a
possible $CP$-violating interaction among gluons in QCD; limits on the
$CP$ violation in strong interactions imply that it must be very small.
The other parameter is associated with the mechanism responsible for the
breakdown of the electroweak $SU(2)\times U(1)$ to $U(1)_{em}$. This can
be taken to be the mass of the as yet undiscovered Higgs boson, whose
couplings are determined once its mass is given.
Additional parameters are needed to accommodate neutrino masses and
mixings.

The gauge theory part of the SM has been well tested, but there is little
direct evidence either for or against the simple Higgs mechanism for
electroweak symmetry breaking. The current experimental lower bound on the
Higgs mass is  $113.5\,\GeV$\cite{leplimit}. If the Standard Model Higgs
sector is correct, then precision measurements at the $Z$ and elsewhere can be
used to constrain the Higgs mass via its contribution to the measured
quantities from higher order quantum corrections to be less than
212~GeV~\cite{lepfits} at 95\% confidence. As the Higgs mass increases, 
its self couplings and its couplings to the $W$ and $Z$ bosons
grow~\cite{Lee-quigg}. This feature has a very important consequence.  Either
the Higgs boson must have a mass less than about 800~GeV or the dynamics
of $WW$
and $ZZ$ interactions with center of mass energies of order 1~TeV will reveal
new structure. It is this simple argument that sets the energy scale that must
be probed to guarantee that an experiment will be able to provide information
on the nature of electroweak symmetry breaking.

The presence of a single elementary scalar boson is distasteful to many
physicists.
If the theory is part of some more fundamental theory, which has some other larger
mass scale (such as the scale of grand unification or the Planck scale), there is
a serious ``fine tuning'' or naturalness problem. 
Radiative corrections to the Higgs boson mass result in
a value that is driven to the larger scale unless some delicate cancellation is 
engineered ($m_0^2-m_1^2\sim M_W^2$ where $m_0$ and $m_1$ are order $10^{15}$ GeV or larger).
There are two ways out of this problem which involve new physics at a 
scale of order 1~TeV.
New strong dynamics could enter that provides the scale of $M_W$, 
or new particles could appear
so that the larger scale is still possible, but the divergences are canceled 
on a much smaller 
scale. It is also possible that there is no higher scale as, for
example in models with extra dimensions. In any of the options, Standard Model, new dynamics or cancellations, 
the energy scale is the same:
something must be discovered on the TeV scale.

Supersymmetry is an appealing concept for which there is, at present, no
experimental evidence~\cite{susy}. It offers the only presently known
mechanism for incorporating gravity into the quantum theory of particle
interactions, and it provides an elegant cancellation mechanism for the
divergences provided that at the electroweak scale the theory is
supersymmetric.  The successes of the Standard Model (such as precision
electroweak predictions) are retained, while avoiding any fine tuning of the
Higgs mass. Some supersymmetric models allow for the unification of gauge
couplings at a high scale and a consequent reduction of the number of
arbitrary parameters.  Supersymmetric models postulate the existence of
superpartners for all the presently observed particles: bosonic superpartners
of fermions (squarks $\squark$ and sleptons $\tilde \ell$), and fermionic
superpartners of bosons (gluinos $\gluino$ and gauginos $\tilde\chi^0_i$,
$\tilde\chi^\pm_i$).  There are also multiple Higgs bosons: $h$, $H$, $A$ and
$H^\pm$.  There is thus a large spectrum of presently unobserved particles,
whose exact masses, couplings and decay chains are calculable in the theory
given certain parameters.  Unfortunately these parameters are unknown.
Nonetheless, if supersymmetry is to have anything to do with electroweak
symmetry breaking, the masses should be in the region 100~GeV -- 1~TeV.

An example of the strong coupling 
scenario is ``technicolor'' or models based on dynamical symmetry
breaking~\cite{technicolor}.
Again, if the mechanism is to have anything to 
do with Electroweak Symmetry breaking we would expect
new states in the region 100~GeV -- 1~TeV; most models predict a large
spectrum.
An elegant implementation of this appealing idea is lacking.
However, all models predict structure in the $WW$ scattering amplitude 
at around 1 TeV center of mass energy. 

There are also other possibilities for new physics that are not 
necessarily related to the scale of electroweak symmetry breaking.
There could be new neutral or charged gauge bosons with mass larger 
than the $Z$ 
and $W$; there could be
new quarks, charged leptons or massive neutrinos; 
or quarks and leptons could turn out not to be elementary objects. It
is even possible that there are extra space time 
dimensions~\cite{Arkani-Hamed:1998}\cite{Randall:1999ee} that have observable
consequences for energies in the TeV mass range.
While we have no definitive expectations for the masses of these objects,
the LHC must be able to search for them over its
available energy range.


\section{The Large Hadron Collider}

\subsection{Machine parameters}

The LHC machine is a proton-proton collider that will be installed in the 26.6
km circumference tunnel formerly 
used by the LEP electron-positron collider at CERN~\cite{lhcbook}.
The 8.4 Tesla dipole magnets --- each 14.2 meters long (magnetic length) --- 
are of the ``2 in 1'' type:
the apertures for both beams have a common mechanical structure and cryostat.
These superconducting 
magnets operate at 1.9K and have an aperture of 56 mm. They will
be placed on the floor in the LEP ring after removal and storage of LEP.
The 1104 dipoles and 736 quadruples support beams of 7~TeV energy and a
circulating current of 0.54 A. 

Bunches of protons separated by 25 ns and with an RMS length of 75 mm
intersect at four points where experiments are placed. Two of these 
are high luminosity regions and house the ATLAS  and CMS detectors. Two other
regions house the ALICE detector~\cite{alice}, to be used for the study of heavy ion
collisions, and LHC-B~\cite{lhcb}, a detector optimized for the study of 
$b$-mesons and $b$-Baryons.
The beams cross at an angle of 200$\mu$rad, resulting in peak luminosity of
$10^{34}\,\cmsec$ with a luminosity-lifetime of 
10 hours. The expected data samples are $30$ ($300$) fb$^{-1}$ at
 $10^{33}\,\cmsec$ ($10^{34}\,\cmsec$), called  low (high) luminosity
 in this document. At the peak luminosity there are an average of $\sim 20 pp$ interactions
per bunch crossing. Ultimately, the peak luminosity may increase beyond
$10^{34}$ cm$^{-2}$ sec$^{-1}$.
The machine will also be able to accelerate heavy
ions resulting in the possibility of, for example,  Pb-Pb collisions at 1150 TeV in the center
of mass and luminosity up to $10^{27}$ cm$^{-2}$ sec$^{-1}$.

In the $pp$ version, which will be the focus of the rest of this article,
the LHC can be thought of as a parton-parton collider with beams of partons of
indefinite energy. The effective luminosity~\cite{ehlq}
of these collisions is proportional to the $pp$ luminosity and falls rapidly with
the center of mass energy of the parton-parton system. The combination of the
higher energy and luminosity of the LHC compared to the highest energy collider
currently operating, the Tevatron, implies that the accessible energy range is
extended by approximately a factor of ten.
 
\subsection{Physics Goals}

The fundamental goal is to uncover and explore the 
physics behind electroweak symmetry breaking. This involves the following specific
challenges:
\begin{itemize}
\item
Discover or exclude the Standard Model Higgs and/or the multiple Higgs
bosons of
supersymmetry.
\item
Discover or exclude supersymmetry over the entire theoretically allowed mass
range up to a few TeV.
\item
Discover or exclude new dynamics at the electroweak scale.
\end{itemize} 
The energy range opened up by the LHC also gives us the opportunity to search
for other, possibly less well motivated, objects:
\begin{itemize}
\item Discover or exclude any new electroweak gauge bosons with masses below 
several TeV.
\item Discover or exclude any new quarks or leptons that are
kinematically accessible.
\item Discover or exclude extra-dimensions for which the appropriate mass
scale is below several TeV.
\end{itemize}
Finally we have the possibility of exploiting the enormous production rates for
certain Standard Model particles to conduct the following studies:
\begin{itemize}
\item Study the  properties of the top quark and set limits on exotic
decays such as $t\to c Z$ or $t\to b H^+$.
\item Study of $b$-physics, particularly that of $b$-baryons and $B_s$ mesons.
\end{itemize}

An LHC experiment must have the ability to find the unexpected. New phenomena of whatever type
will decay into the particles of the Standard Model. In order to cover the lists given above
a detector must have great flexibility. The varied physics signatures for these processes 
require the ability to reconstruct and measure final states involving the following
\begin{itemize}
\item Charged leptons, including the $\tau$ via its hadronic decays.
\item The electroweak gauge bosons: $W$, $Z$ and $\gamma$.
\item Jets of energy up to several TeV coming from the production at high transverse momentum of quarks 
and gluons.
\item Jets that have $b$-quarks within them.
\item Missing transverse energy carried off by weakly interacting neutral 
particles such as neutrinos. 
\end{itemize}

In the discussion of physics signals that we present below, it is necessary to
estimate production cross sections for both signal and background processes.
This is done using perturbative QCD. Such calculations depend on the parton
distribution functions that are used, the energy ($Q^2$ scale) used in the
evaluation of the QCD coupling constant and the structure functions, and the
order in QCD perturbation theory that is used. These issues make comparison
between different simulations of the same process difficult. Higher order
corrections are not known for all processes and in some cases they are known
for the signal and not for the background. Most of the LHC simulations are conservative and 
use lowest order
calculations. Higher order corrections almost always increase the cross
sections, typically by a so-called $K$ factor of order 1.5 to 2.0. The real
analysis will of course be based on the best calculations available at the
time. At present, the uncertainties from the choice of scale and structure
functions are typically at the 20\% level.
 The total cross-section for $b$-quark production is
particularly uncertain.

The level of simulation used to study  the processes varies quite widely.
For some processes a full GEANT~\cite{geant} style simulation has 
been carried out.
Such simulations are very slow 
($\sim 10$~Spec95-hr/event)
and are difficult to carry out for processes where a large number of
events needs to be simulated and many strategies for extracting signals 
need to be pursued. In these cases a particle level simulation and
parameterized detector response is used. A lower level of simulation
involving partons ({\it i.e.}, leptons and jets) and parameterized response
is fast  and might be required when the underlying parton 
process is not present in full event generators. This last level of
simulation is useful for exploring signals but often leads to
overly optimistic results, particularly when the reconstruction
of invariant masses of jets or missing energy are involved. None of the results
included here use this last level of simulation, unless stated explicitly.   

\subsection{Detectors}

Two large, general-purpose $pp$ collider detectors are being constructed for
LHC:  ATLAS~\cite{atlas} and CMS~\cite{cms}.  
Both collaborations completed Technical Proposals for
their detectors in December 1994, and were formally approved
in January 1996. Construction is now underway.  Though they differ in most details, the detectors 
share many common features that are  derived from the physics goals of LHC:
\begin{itemize}
\item they both include precision electromagnetic calorimetry;
\item they both use large magnet systems (though of different geometries)
in order to obtain good muon identification and precision momentum 
measurement;
\item they both have lepton identification and precision measurement over $|\eta| < 2.5$;
\item
they both have  multi-layer silicon pixel
tracker systems for heavy flavor tagging (the usefulness of this capability is
an important lesson from the Tevatron); 
\item
they both include  calorimetry  for large $|\eta| < 5$ coverage in order to
obtain the required $\met$ resolution.
\end{itemize}

The ATLAS detector is shown uses a
tracking system employing silicon pixels, silicon strip detectors, and a
transition radiation tracker, all contained within a 2~Tesla
superconducting solenoid.  The charged track resolution is $\Delta
p_T/p_T=20\% $ at $p_T=500\,$GeV.  The tracker is surrounded by an
electromagnetic calorimeter using a lead-liquid argon accordion design;
the EM calorimeter covers $|\eta|<3$ (with trigger coverage of
$|\eta|<2.5$) and has a resolution of $\Delta E/E = 10\%/\sqrt{E} \oplus
0.7\%$.  The hadronic calorimeter uses scintillator tiles in the barrel
and liquid argon in the endcaps ($|\eta|>1.5$); its resolution is
$\Delta E/E = 50\%/\sqrt{E} \oplus 3\%$.  Forward calorimeters cover the
region $3 < |\eta| < 5$ with a resolution better than $\Delta E/E =
100\%/\sqrt{E} \oplus 10\%$.  Surrounding the calorimeters is the muon
system.  Muon trajectories are measured using three stations of
precision chambers (MDT's and CSC's) in a spectrometer with bending
provided by large air-core toroid magnets. The resulting muon momentum
resolution is $\Delta p_T/p_T=8\%$ at $p_T=1\,$TeV and $\Delta
p_T/p_T=2\%$ at $p_T=100\,$GeV.  Muons can be triggered on over the
range $|\eta|<2.2$.

The CMS detector is has
calorimeters and tracking system
are contained in a 4~Tesla superconducting coil which provides 
the magnetic field
for charged particle tracking.  
The tracking system is based on silicon pixels and silicon strip
detectors.
The charged track resolution is $\Delta p_T/p_T=5\%$ at $p_T=1\,$TeV
and $\Delta p_T/p_T=1\%$ at $p_T=100\,$GeV.
CMS has chosen a precision electromagnetic calorimeter 
using lead tungstate (PbW0$_4$) crystals, covering 
$|\eta|<3$ (with trigger coverage of $|\eta|<2.6$).
Its resolution at low luminosity
is $\Delta E/E = 3\%/\sqrt{E} \oplus 0.5\%$.  The surrounding
hadronic calorimeter uses scintillator tiles in the barrel and
endcaps; its resolution for jets (in combination with the
electromagnetic calorimeter) is $\Delta E/E = 110\%/\sqrt{E} \oplus 5\%$.  The
region $3 < |\eta| < 5$ is covered by 
forward calorimeters using parallel-plate chambers or quartz fibers 
and having a resolution of about 
$\Delta E/E = 180\%/\sqrt{E} \oplus 10\%$.
Muon trajectories outside the coil are measured in four layers
of chambers (drift tubes and CSC's) embedded in the iron return yoke.  
The muon momentum measurement using the muon chambers and the
central tracker covers the range 
$|\eta|<2.4$ with a resolution $\Delta p_T/p_T=5\%$ at
$p_T=1\,$TeV and  $\Delta p_T/p_T=1\%$ at $p_T=100\,$GeV.  The muon
trigger extends over $|\eta|<2.1$.

\section{Higgs Physics}

We will use ``Higgs bosons'' to refer to any scalar particles whose
existence is connected to electroweak symmetry breaking. Generically,
Higgs bosons couple most strongly to heavy particles.  Their production
cross section in hadron colliders is small compared to QCD
backgrounds,
 resulting in final states
with low rates or low signal-to-background ratios. The ability to detect
them and measure their mass provides a set of benchmarks by which
detectors can be judged.  A specific model is required in order to
address the quantitative questions of how well the detector can perform.
While one may not believe in the details of any particular model, a
survey of them will enable general statements to be made about the
potential of the LHC and its detectors.

\subsection{Standard Model Higgs}

\begin{figure}
\dofig{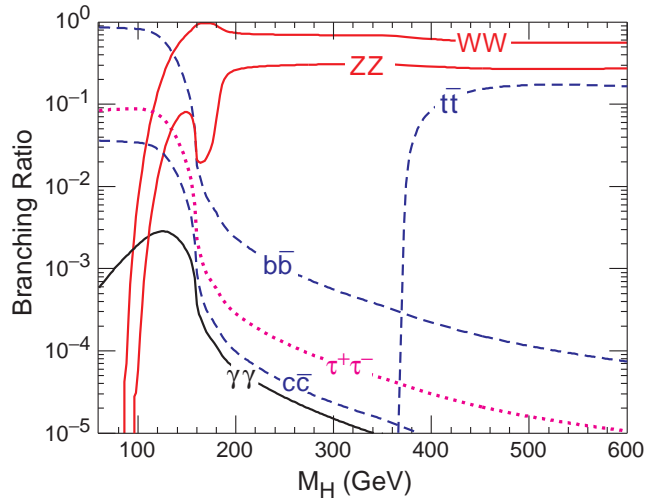}
\caption[]{The branching ratios of the 
Standard Model Higgs boson as a function of its mass.
\label{Higgsbr}}
\end{figure}

All the properties of the Standard Model Higgs boson are determined once its
mass is known; the search strategy at LHC is therefore well defined.  The
current limit on the Higgs mass from experiments at LEP~\cite{leplimit} is
$M_H>113.5$ GeV.  There are several relevant production mechanisms at LHC;
$gg\to H$ via a heavy quark loop; $q\overline{q}\to WH$; $gg\to
t\overline{t}H$; $gg\to b\overline{b}H$ and $qq\to qqH$  (``$WW$
fusion'').
  The relative
importance of these processes depends upon the Higgs mass, the first dominates
at small mass and the two become comparable for a Higgs mass of 1 TeV.
The Higgs branching ratios are
shown in Fig.~\ref{Higgsbr}.

\subsubsection{$H \to \gamma \gamma$ and associated production channels} 

At masses just above the range probed by LEP, the dominant decay of the Higgs
boson is to $b\overline{b}$, which is essentially impossible to separate from
the huge QCD $b\overline{b}$ background. The decay to $\gamma\gamma$ is the
most promising channel in this region. The branching ratio is very small, and
there is a large background from the pair production of photons via
$q\overline{q}\to \gamma\gamma$, $gg \to \gamma\gamma$, and the bremsstrahlung
process $qg \to q(\to\gamma)\gamma$. Excellent photon energy resolution is
required to observe this signal. Hence, this process is one that drives the
very high quality electromagnetic calorimetry of both ATLAS and CMS.

CMS has a mass resolution of 540~(870)~MeV at $m_H=110\,\GeV$ for low (high)
luminosity~\cite{cmsgamgam}.  The mass resolution is worse at high luminosity
due to event pile up.
 The ATLAS mass resolution at low (high) luminosity is
1.1~(1.3)~GeV for  $M_H=110$ GeV. The photon acceptance and identification
efficiency are higher in the ATLAS analysis~\cite{atlasgamgam}, partly because
CMS rejects some of the photons that convert in the inner detector.

\begin{figure}[t]
\dofigf{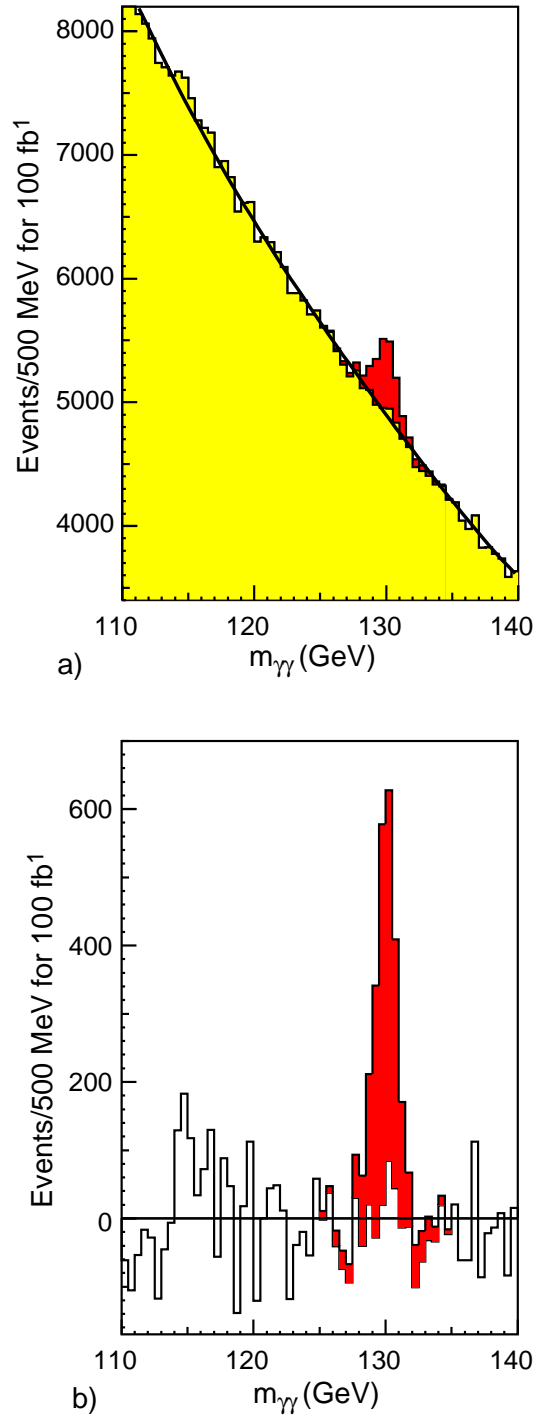}
\caption[]{(a) The invariant mass distribution of $\gamma\gamma$ pairs for
$M_h=130\,\GeV$ as simulated by the CMS collaboration. (b) Same, with a smooth
background fitted and subtracted. From Ref.~\cite{cmsecal}.
\label{fighiggam}}
\end{figure}
 
In addition to the background from $\gamma\gamma$ final states, there are
$\jet-\gamma$ and $\jet-\jet$ final states, that are much larger. A
$\jet/\gamma$ rejection factor of $\sim 10^3$ is needed to bring these
backgrounds below the irreducible $\gamma\gamma$ background.  A detailed GEANT
based study of the ATLAS detector has been performed for  these
backgrounds~\cite{atlasgamgam}.  Jets were rejected by applying cuts on the
hadronic leakage,  isolation and the measured width of the
electromagnetic shower.  These cuts result in an estimate of these backgrounds
which is a factor of four below the irreducible $\gamma\gamma$ background.  
 There are uncertainties in the
rates for these ``reducible'' backgrounds, however  one can be confident that
they are smaller after cuts than the irreducible $\gamma\gamma$ background.

In the CMS analysis for this process~\cite{cms,cmsgamgam}, two isolated
photons are required, one with $p_T>25$ GeV and the other with $p_T>40$ GeV.
Both are required to satisfy $\abseta <2.5$ and to have no track or additional
electromagnetic energy cluster with $p_T>2.5\,\GeV$ in a cone of size $\Delta
R=0.3$ around the photon direction.  The Higgs signal then appears as a peak
over the smooth background.  The signal-to-background ratio is small, but
there are many events. A curve can be fitted to the smooth background and
subtracted from data.  Fig.~4 shows the total and background-subtracted
distributions for a Higgs mass of $130\,\GeV$. For an integrated luminosity of
$100\,\fbi$ it is possible to discover a Higgs using this mode if its mass is
between the lower limit set by LEP and about $140\,\GeV$. A signal can also be
found over a more limited mass range for an integrated luminosity of
$10\,\fbi$. Results of the ATLAS study are similar~\cite{atlasgamgam}.

Another process is available at the lower end of the mass range. If the Higgs is produced in association with a $W$ or
$t\overline{t}$, the cross section is substantially reduced, but the
presence of additional particles provide a  proportionally larger
reduction in the background.  Events are required to have an isolated lepton arising from the decay of the $W$ (or top quark).
This lepton  can be used to determine the vertex
position.
The process is only useful at high luminosity as, for $10$ $\fbi$, 
there are approximately 15 signal events for Higgs masses between 90 and 120 GeV (the falling cross-section is compensated by the
increased branching ratio for $H\to \gamma\gamma$)
over an approximately equal background~\cite{AtlasPhysTDR}. 
The process will therefore provide confirmation of a discovery
made in the $\gamma\gamma$ final state without an associated lepton
and measurements of the couplings.

\subsubsection{$H \to b \overline{b}$}

The dominant decay of a Higgs boson if its mass is below $2M_W$ is to
$b\overline{b}$. The signal for a Higgs boson produced in isolation is
impossible to extract: there is no trigger for the process and the background
production of $b\overline{b}$ pairs is enormous. The production of a Higgs
boson in association with a $W$ or $t\overline{t}$ pair can provide a high
$p_T$ lepton that can be used as a trigger. A study was conducted by ATLAS of
this very challenging channel (see Section of 19.2.4 of
Ref.~\cite{AtlasPhysTDR}).  Events were triggered by requiring a muon
(electron) with $\abseta <2.5$ and $p_T>6 (20)$ GeV.  

\begin{figure}[t]
\dofig{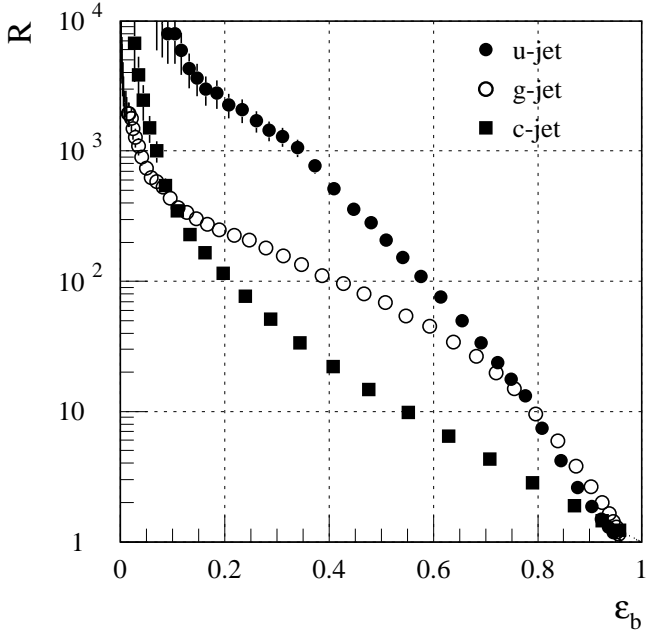}
\caption[]{Rejection factor for jets produced from $u$ and $c$ quarks and
gluons  at low luminosity as a function of the tagging efficiency for 
$b$-quark jets in the ATLAS
detector. Processes such as $g \to b \overline b$ are
included as mistags. From Ref.~\cite{AtlasPhysTDR}.
\label{atlasbtag}}
\end{figure}

\begin{figure}
\dofig{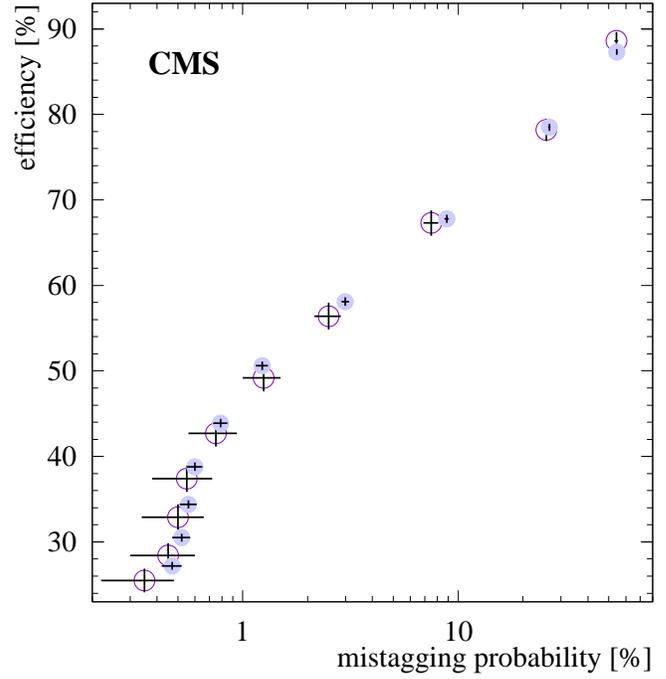}
\caption[]{Mistagging probability  for jets produced from $u$  quarks 
 as a function of the tagging efficiency for $b$-quark jets in the CMS
detector with the all silicon tracker. From Ref.~\cite{cmstag}.
\label{cmsbtag}}
\end{figure}

The  expected $b$-tagging efficiency for ATLAS 
was determined by full simulation of
samples of $H \to b\overline{b}$, $H \to u \overline{u}$, and $t\overline{t}$
events. The results from these samples for the $b$-tagging rate and rate of
fake tags from non $b$-jets were obtained over a range of $p_T$.
 The results can therefore be extrapolated to other
cases --- e.g,
$b$-jets in supersymmetry
events --- that have not been fully simulated. The ATLAS detector has a
pixel layer 
at $\sim 5$~cm from the beam.  The $b$-tagging efficiency
is correlated with the rejection factor that is obtained against other jets as
is shown in Fig.~\ref{atlasbtag}.  The rejection of charm jets is limited by
the lifetime of charged hadrons and that of gluons by the production of
$b\overline{b}$ pairs in the jet itself. Note that rejection factors $\sim
100$ against light quark jets can be obtained for a $b$-tagging efficiency of
60\%. The $b$-tagging efficiency for CMS has similarly
been determined from full simulation and is shown in
Figure~\ref{cmsbtag}. As in the case of ATLAS the pixel layers at
radii of  4, 7, and 11 cm are used for the tagging.
These $b$-tagging efficiencies  are not significantly
different from those already obtained by CDF~\cite{cdftag}.

\begin{figure}
\dofig{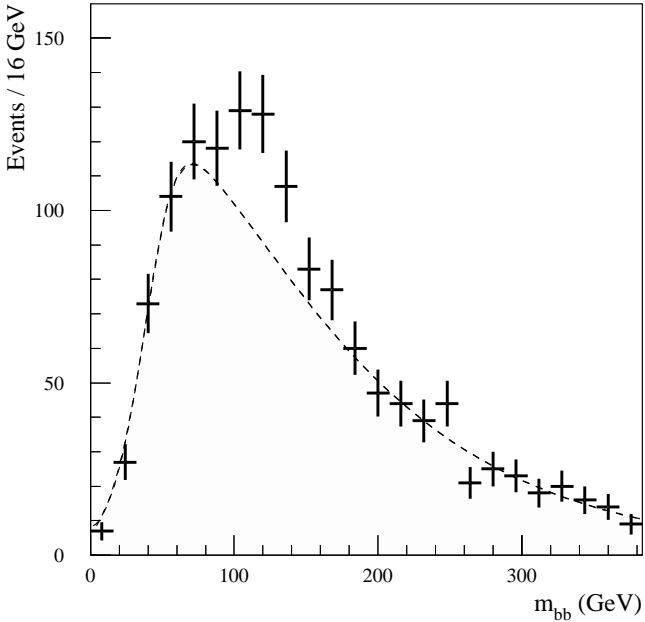}
\caption[]{Invariant mass distribution, $m_{b\overline b}$, of tagged $b$-jet
pairs in fully reconstructed $t\overline{t}H$ signal events with a Higgs-boson
mass of 120 GeV above the summed background (see text), for an integrated
luminosity of $100\,\fbi$ ($30\,\fbi$ with low-luminosity operation and
$70\,\fbi$ with high-luminosity operation). The points with error bars
represent the result of a single experiment and the dashed line represents the
background distribution. From Ref.~\cite{AtlasPhysTDR}.
\label{Atlashtobb}}
\end{figure}

The ATLAS study of $H\to b\overline{b}$ uses the tagging efficiencies from
the full simulation study. Both the $t\overline{t}H$ and $WH$ final
states were studied but the former is more powerful, so it is the only
one discussed here.
Jets were retained if they 
had $p_T>15$ GeV. This threshold was raised to 30 GeV for simulations
at a luminosity of 
$10^{34}$ cm$^{-2}$ sec$^{-1}$.
In order to reduce  the
background  a veto was applied to reject events 
with a second isolated lepton $p_T>6$ GeV and $\abseta
<2.5$ and a total of 4 tagged $b$-jets was required. 
Reconstruction of both top quarks using a kinematic fit is essential
to reduce the combinatorial $b\overline{b}$ background. For a 
luminosity of $100$ $\fbi$, there are  107 and 62 signal
events for Higgs masses of  100 and 120 GeV. 
The reconstructed $b\overline{b}$ mass distribution is approximately Gaussian
with $\sigma/M \sim 0.2$; 
it has a tail on the low side caused mainly by 
gluon radiation off the final state $b$ quarks and losses due to decays. 
The background arising from $t\overline{t}jj $ events is the most important;
approximately 250 events in a 
bin of width 30 GeV centered on the reconstructed $b\overline{b}$ mass
peak.   Fig.~\ref{Atlashtobb} shows the reconstructed
$b\overline{b}$ mass distributions for the summed signal and
background for $m_H=120$ GeV.
 Extraction of a signal will be possible if at all
only over a   limited mass range ($\sim 80-120$~GeV) and depends 
critically upon the $b$-tagging efficiency and background rejection.
The signal will provide a second observation of the  Higgs boson in
this mass range 
 and also provide valuable information on the Higgs
couplings.

\begin{figure}
\dofig{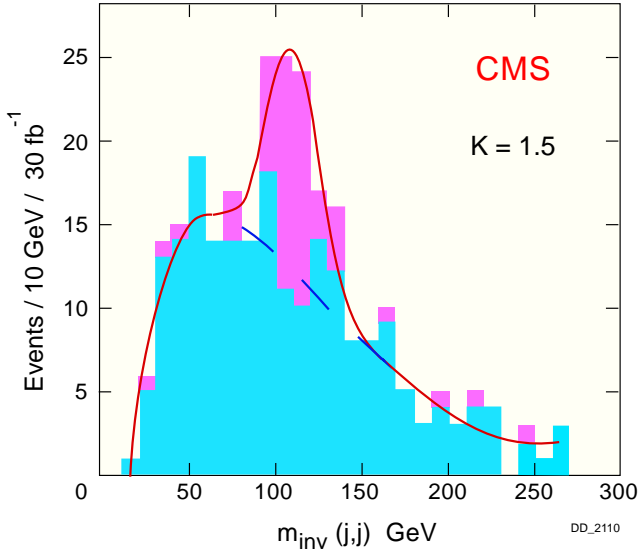}
\caption[]{Reconstructed mass distribution, $m_{j,j}$, of tagged $b$-jet
pairs showing a   $t\overline{t}H$ signal  with a Higgs-boson
mass of 115  GeV above the background, for an integrated
luminosity of  $30\,\fbi$.  From Ref.~\cite{drollinger}.
\label{cmstth}}
\end{figure}

A similar analysis has been performed by CMS \cite{drollinger}.
Events were required to have an isolated $e$ or $\mu$ with $p_T>10
\GeV$, six jets with $p_T>20
\GeV$, four of which are tagged as $b$'s. A K-factor of 1.5 is
included for the signal only. A likelihood analysis gave the results
shown in Fig.~\ref{cmstth} with $S/B= 0.73$. The extraction of the
$t\overline{t}H$ Yukawa coupling from this signal was also studied. 

\subsubsection{$H \to ZZ^* \to 4\ell$}

The search for the Standard Model Higgs relies on the four-lepton channel over
a broad mass range from $m_H \sim 130\,$GeV to $m_H \sim 800\,$GeV.
Below $2m_Z$, the event rate is small and the background reduction more
difficult, as one or both of the $Z$-bosons are off-shell.  In this mass region
the Higgs width is small ($\simle 1\,$GeV) and so lepton energy or momentum
resolution is of great importance in determining the significance of a 
signal~\cite{atlaszzstar}.

For $m_H < 2m_Z$, the main backgrounds arise from $t\overline t$, $Zb\overline
b$ and continuum $ZZ^*/Z\gamma^*$ production.  Of these, the $t\overline t$ 
background
can be reduced by lepton isolation and by lepton pair invariant mass cuts.
The $Zb\overline b$ background cannot be reduced by a lepton pair invariant
mass cut but can be suppressed by isolation requirements and impact
parameter cuts.  The $ZZ^*$
process is an irreducible background.
Both CMS and ATLAS studied the process for $m_H = 130$, 150 and 170~GeV.  
Signal events
were obtained from both $gg \to H$ and $WW/ZZ$ fusion processes, giving
consistent cross sections 
$\sigma\cdot B \approx 3$, 5.5 and 1.4~fb respectively (no $K$-factors being
included). 

\begin{figure}[t]
\dofigf{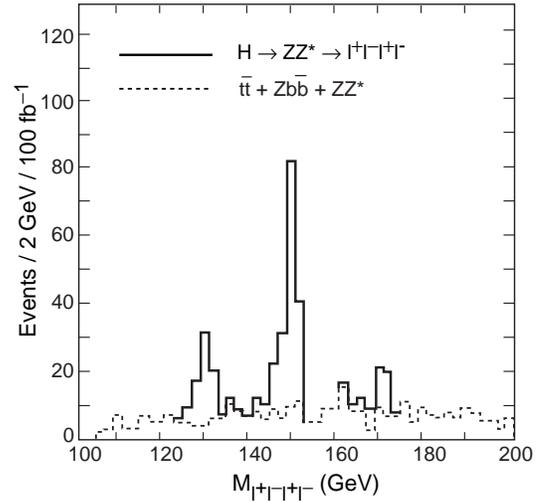}
\vskip10pt
\dofigf{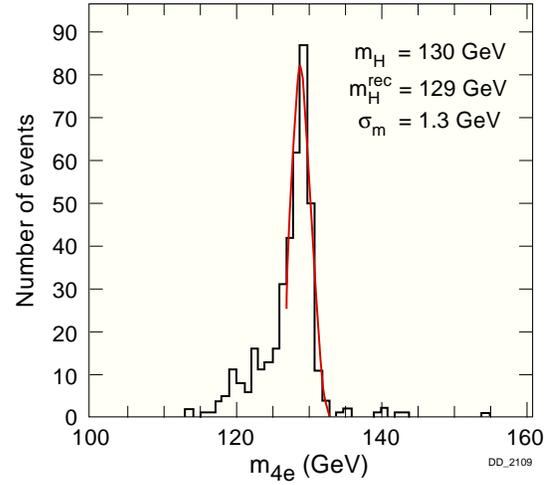}
\caption[]{Top: Reconstructed four-lepton mass in CMS for $H\to 4\ell$ for
various masses and sum of all backgrounds with $100\,\fbi$.
Bottom: Reconstructed four-electron mass 
for $m_H = 130$ GeV,  showing the radiative tail predicted by full
simulation. From Ref.~\protect\cite{cms4lres}.
\label{figzzstar}}
\end{figure}

In the CMS study~\cite{cms,cmszzstar}
event pileup appropriate to 
${\cal L}=10^{34}\,{\rm cm}^{-2}{\rm s}^{-1}$ was modeled by superimposing 15
minimum bias events (simulated by QCD dijets with $p_T \geq 5\,$GeV).
The muon resolution was obtained from a full simulation of the detector
response and track-fitting procedure.  
This was then parameterized as a function
of $p_T$ and $\eta$.  Internal bremsstrahlung was generated using the PHOTOS
program and leads to about 8\% of reconstructed $Z \to \mu^+ \mu^-$ pairs 
falling outside a $m_Z \pm 2 \sigma_Z$ window for $m_H = 150\,$GeV.  
The reconstructed $\mu^+ \mu^-$ mass has a resolution $\sigma_Z = 1.8\,$GeV in
the Gaussian part of the peak.  
The electron response was obtained from a full GEANT simulation of the
calorimeter, including the effects of material in the beampipe and the tracker,
and the reconstruction of electron energy in the crystal calorimeter.
Including internal and external bremsstrahlung, and using a $5\times 7$ crystal
matrix to reconstruct the electron, the mass resolution $\sigma_Z = 2.5\,$GeV
and the reconstruction efficiency is about 70\% (within $m_Z \pm 2 \sigma_Z$).

Events were selected which had one electron with $p_T > 20\,$GeV, one with
$p_T > 15\,$GeV and the remaining two with $p_T > 10\,$GeV, all within
$|\eta| < 2.5$.  For muons, the momenta were required to exceed 20, 10 and
5~GeV within $|\eta| < 2.4$.  One of the $e^+e^-$ or $\mu^+\mu^-$ pairs was
required to be within $\pm 2\sigma_Z$ of the $Z$ mass.  This cut loses
that fraction of the signal where both $Z$'s are off-shell, about a 24\%
inefficiency at $m_H = 130\,$GeV and 12\% at $m_H = 170\,$GeV.
The two softer leptons were also required to satisfy $m_{\ell\ell} > 12\,$GeV.
Additional rejection is obtained by requiring 
that any three of the four leptons be isolated in the tracker, demanding that
there is no track with $p_T > 2.5\,$GeV within the cone $R<0.2$ around the
lepton.  This requirement is not very sensitive to pileup as the $2.5\,$GeV
threshold is quite high.  This yields signals at the 
level of 7.4, 15.2 and 5.0 standard deviations for
$m_H = 130$, 150, and 170~GeV in $200 \, \fbi$.  
The four-lepton mass distributions are shown in Fig.~\ref{figzzstar}
which also shows the $4e$ final state. The latter clearly shows the
effect of bremsstrahlung.

The ATLAS~\cite{AtlasPhysTDR} study followed a similar technique.  
The detector resolutions and reconstruction efficiencies were obtained using
detailed detector simulations, including the effects of pileup. 
Events were selected which had two leptons with $p_T > 20\,$GeV, and 
the remaining two with $p_T > 7\,$GeV, all within
$|\eta| < 2.5$.  One of the $e^+e^-$ or $\mu^+\mu^-$ pairs was
required to be within $\pm m_{12},$GeV of the $Z$ mass.  
The two softer leptons were also required to satisfy 
$m_{\ell\ell} > m_{34}$. $m_{12}$ and $m_{34}$ are varied as a
function of the Higgs mass; for $m_H=130$ GeV,  $m_{12}=10$ GeV and
$m_{34}=30$ GeV. 
For the four-electron mode, the Higgs mass resolution at $m_H = 130\,$GeV
is 1.8~(1.5)~GeV at high (low) luminosity, including the effect of electronic 
noise in the calorimeter.  For muons, the
corresponding figure is 1.4~GeV after correcting for muon energy losses in the
calorimeter and  combining the muon
momentum measured in the muon system with that obtained from the central
tracker after the tracks have been matched.

ATLAS used a combination of calorimeter isolation and impact parameter
cuts to reject background from $Zb\overline{b}$  and $t\overline{t}$ events.
  The isolation criterion is that the transverse energy
within $R=0.2$ of the lepton be less than $E_T^{cut}$ or 
that there  are no additional reconstructed tracks above a threshold
in the cone. The rejections obtained by these methods are correlated.
 Values of $E_T^{cut}$
of 3, 5, and 7~GeV were used for $4\mu$, $ee\mu\mu$ and $4e$ modes at
$10^{33}$
($10^{34}$) luminosity to obtain a constant signal efficiency of 85\% (50\%).
Tighter cuts can be used for muons because they do not suffer from
transverse leakage of the EM shower.  
The impact parameter, as measured in
the pixel layers, is used to further reduce the background from heavy flavor
processes ($t \overline t$ and $Z b \overline b$)~\cite{AtlasPhysTDR}.  
ATLAS obtain signals at the 
level of 10.3 (7.0), 22.6 (15.5) and 6.5 (4.3) standard deviations for
$m_H = 130$, 150, and 170~GeV in 
$100\,\fbi$ $(30\, \fbi)$.

\subsubsection{$H \to WW^{(*)} \to \ell^+\nu\ell^-\bar\nu$}

The decay $H \to WW^{(*)} \to \ell^+\nu\ell^-\bar\nu$ can provide
valuable information in the mass region around 170~GeV where the
branching ratio $H \to 4\ell$ is reduced~\cite{Dittmar:1997ss}. For this
mass the two-body $WW$ decay dominates, so $BR(H \to WW^{(*)} \to
\ell^+\nu\ell^-\overline\nu) / BR(H \to 4\ell)\sim 100$.

\begin{figure}[t]
\dofig{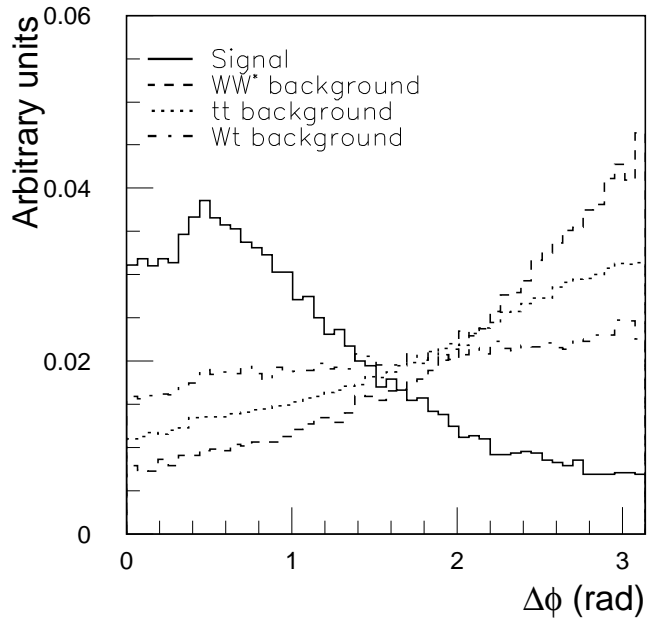}
\caption[]{Difference in azimuth between the two leptons for $H \to WW^*
\to\ell\nu\ell\nu$ signal events with $m_H = 170$ GeV and for the
$WW^*$, $t\overline{t}$ and $Wt$ background events. All distributions
are normalized to unity.From
Ref.~\cite{AtlasPhysTDR}.
\label{atlaswwphi}}
\end{figure}

For the $\ell^+\nu\ell^-\overline\nu$ final state, the Higgs mass cannot be
reconstructed, so the signal must be observed from an excess of events.  The
dominant background arises from the production of $W$ pairs after cuts to
remove the $t\overline{t}$ background. The ATLAS analysis~\cite{AtlasPhysTDR}
requires:
\begin{itemize}
\item Two isolated opposite sign leptons with $\abseta <2.5$ and
$p_T>20,10$ GeV. In addition the pair must satisfy $M_{\ell\ell}<80\,\GeV$, $\Delta\phi_{\ell\ell} <1$, and
$\Delta\eta_{\ell\ell} <1.5$.
\item No jets with $p_T>15$ GeV and $\abseta <3.1$.
\item $\Etmiss > 40\,\GeV$.
\item A $\ell\ell\Etmiss$ transverse mass between
$m_H-30\,\GeV$ and $m_H$.  
\end{itemize}
At luminosity of $10^{34}$ cm$^{-2}$ sec$^{-1}$, the jet veto is raised
to $30\,\GeV$. After these cuts the signal to background ratio is
approximately 2:1 and there are 340 signal events for $m_H=170\,\GeV$
for 30 $\fbi$. The
signal can be clearly established by looking at the distribution in the
azimuthal separation of the leptons ($\Delta\phi$). As is shown in
Fig.~\ref{atlaswwphi}, this is peaked at small (large) values of
$\Delta\phi$ for the signal (background).

\begin{figure}[t]
\dofig{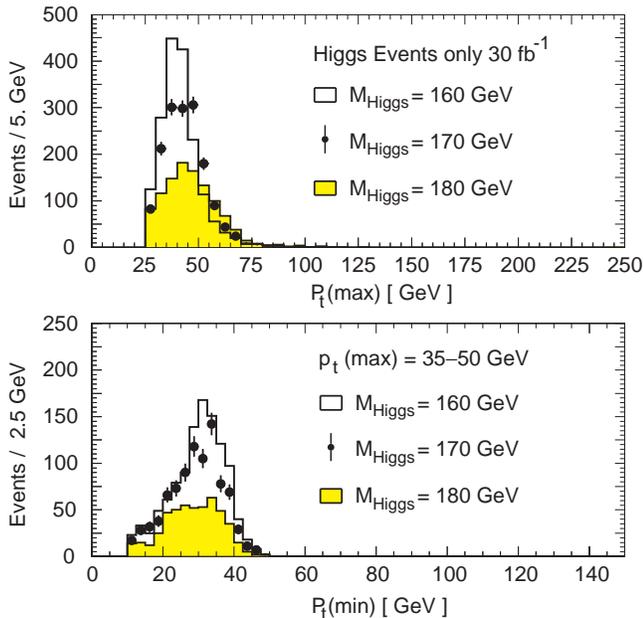}
\caption{Dependence of the lepton kinematics for the $H \to WW \to
\ell^+\nu\ell^-\bar\nu$ signal on the Higgs mass.  From
Ref.~\cite{Dittmar:1998nh}. \label{cmsh2l}}
\end{figure}

Some information on the Higgs mass can be obtained from the lepton
kinematics. This is shown in Fig.~\ref{cmsh2l} from a CMS analysis. This
figure shows the distribution of the larger and smaller $p_T$'s of the
two leptons for an assumed mass of $170\,\GeV$ in comparison with the
expectation for masses of $160\,\GeV$ and $180\,\GeV$. 

As the Higgs mass falls significantly below $150\,\GeV$ the event rate becomes
small. The observability of the signal depends crucially on the ability to
correctly predict the background. 
This background estimation can be checked by comparison with the $ZZ$
final state and by measuring the $WW$ system away from the signal
region. A 5\% 
systematic error on
the background  can be expected;
then a $5\sigma$ observation can be made in the range $130
\GeV <m_H<190\ GeV$ for 30 $\fbi$ of integrated luminosity
\cite{AtlasPhysTDR},

\subsubsection{$H \to ZZ \to 4\ell$}

The $H \to ZZ \to 4\ell$ channel is sensitive over a wide range of Higgs
masses from $2m_Z$ upwards: to about 400~GeV with $10\,\fbi$ and
to about 600~GeV with $100\,\fbi$.  For lower Higgs masses, the width
is quite small and precision lepton energy and momentum measurements are
helpful; for larger masses the natural Higgs width becomes large.
The main background is continuum $ZZ$ production.

\begin{figure}
\dofig{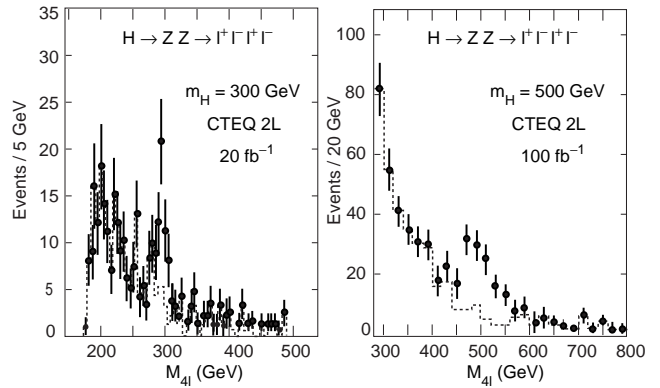}
\caption[]{Mass distribution in $H\to ZZ\to 4\ell$ for various values of $m_H$
as simulated by CMS including all bremsstrahlung losses.  From
Ref.~\cite{Bomestar}. \label{figzz}}
\end{figure}

CMS~\cite{cms,cmszzstar} studied the process for $m_H = 300$, 500 and 600~GeV.  
The electron and muon resolutions and the selection cuts were the same as used
for the $ZZ^*$ channel.   Two  $e^+e^-$ or
$\mu^+\mu^-$ pairs with a mass within $\pm 6\,$GeV of $m_Z$ were required.
No isolation cut was imposed as the remaining backgrounds are small.
The resulting 4-lepton invariant mass distributions are shown in Fig.~\ref{figzz}.
With $100 \fbi$ a signal in excess of six standard
deviations is visible over the entire range $200 < m_H < 600\,$GeV.  
ATLAS obtains very similar results~\cite{AtlasPhysTDR}.  

\subsubsection{$M_H \sim 1$~TeV ($H \to \ell\ell\nu\nu$, $\ell\ell jj$, 
$\ell\nu jj$)}

As the Higgs mass is increased further, its width increases and its
production rate falls, so one must turn to decay channels that have a
larger branching ratio. The first of these is $H\to
ZZ\to\ell\ell\nu\overline{\nu}$.  Here the signal involves a
$Z$ decaying to lepton pairs and a large amount of missing energy. The
signal appears as a Jacobian peak in the missing $E_T$ spectrum.  There
are more potentially important sources of background in this channel
than in the $4\ell$ final state. In addition to the irreducible
background from $ZZ$ final states, one has to worry about $Z+\jets$
events where the missing $E_T$ arises from neutrinos in the jets or from
cracks and other detector effects that cause jet energies to be
mismeasured. At high luminosity the background from the pile up of
minimum bias events produces a $\Etmiss$ spectrum that falls very
rapidly and is small for $\Etmiss>100$ GeV, provided the
calorimeter extends to $\abseta < 5$. ATLAS conducted ~\cite{bosman} a
full GEANT based study of this background for which 5000 high transverse
momentum $Z+\jet$ events were fully simulated.  The events were selected
so that a large fraction of them had jets going into the region
$0.9<\abseta <1.3$ where ATLAS has weaker jet energy resolution due to
the crack between the endcap and barrel hadron calorimeters.  The
dominant part of the $Z+jets$ background that remains is that where the
missing $E_T$ arises from the semi-leptonic decays of $b$-quarks in the
jets. The contribution from detector effects is not dominant.

\begin{figure}[t]
\dofig{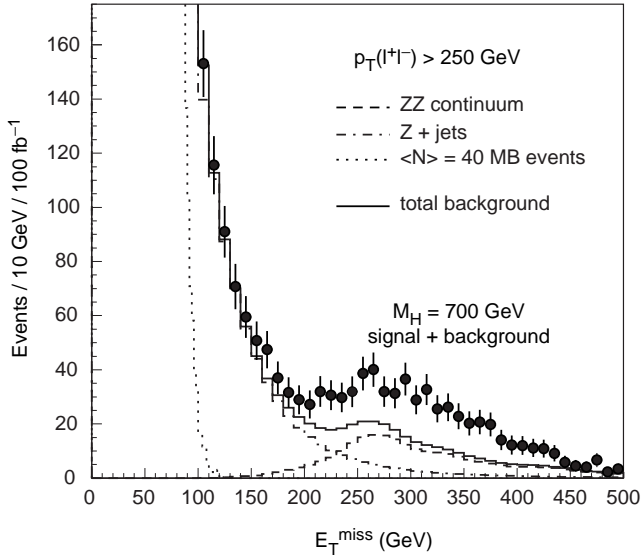}
\caption[]{Missing $E_T$ spectrum for the $H\to ZZ\to \ell\ell\nu\overline{\nu}$ process. The background contributions are shown
separately; $Z+jets$ (dot-dashed); ZZ (dotted) and minimum 
bias pile up (dashed). The signal is due to a Higgs boson of mass 700 GeV.
From Ref.~\cite{atlas}. \label{Higgsllnunu}}
\end{figure}

Fig.~\ref{Higgsllnunu} shows the missing $E_T$ spectrum at high
luminosity ($100\,\fbi$).  On this plot the $Z+{\rm jets}$ background is
estimated from a parton level simulation; there were insufficient
statistics in the full study to obtain this spectrum. This estimate
correctly models the contribution from $b$-decays that the full study
showed to be dominant.  The reconstructed $Z\to \ell\ell$ was required
to have $p_T(Z)>250\,\GeV$; this causes the $ZZ$ background to peak.
(This effect is less pronounced if a cut is made on $\Etmiss$ and then
the plot is remade with $p_T(Z)$ on the abscissa.)  The
dominant $ZZ$ background has QCD corrections of order 40\%~\cite{baur}.
Once data are available, this background will be measured. It signal
to background ratio can be improved significantly by requiring one or
two forward jets at the cost of a smaller acceptance \cite{AtlasPhysTDR}.

The CMS analysis of this process~\cite{cms,cmsllnunu} uses a central jet
veto, requiring that there be no jets with $E_T>150$ GeV within
$\abseta<2.4$.  By requiring a jet in the far forward region (see
below), most of the remaining $ZZ$ background can be rejected. A study
by CMS requiring a jet with $E>1\,\TeV$ and $2.4 < |\eta| <4.7$, produces
an improvement of approximately a factor of three in the signal to
background ratio at the cost of some signal. This mode is only effective
for high mass Higgs bosons and becomes powerful only at high luminosity.
Nevertheless it will provide an unambiguous signal.

\begin{figure}
\dofig{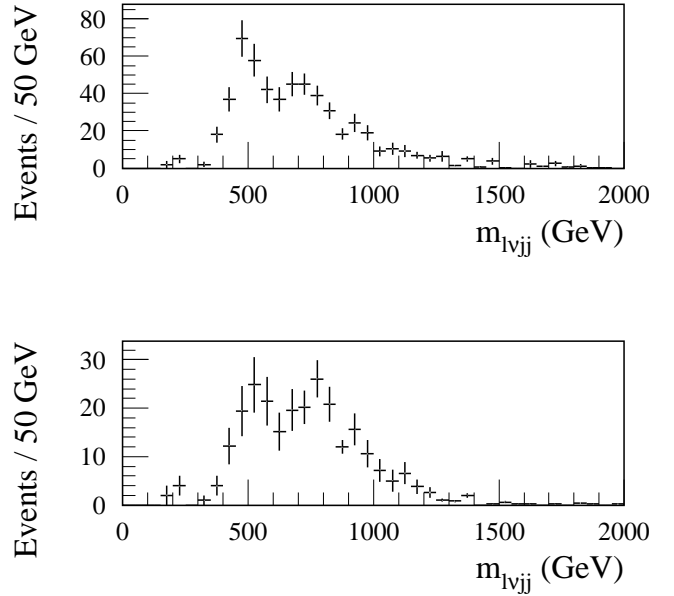}
\caption[]{For an integrated luminosity of 30 $\fbi$ and for $m_H = 800$
  GeV, distribution of $M_{l\nu jj}$ in ATLAS 
for the summed signal + background after
requiring two tag jets with $E_{\rm tag} > 200\,\GeV$ (top) and 
$E_{\rm tag} > 400\,\GeV$ (bottom). From Ref.~\cite{AtlasPhysTDR}.
\label{atlashww}}
\end{figure}

Substantially larger event samples are available if the decay 
modes $H\to WW\to \ell\nu +\jets$ and $H\to ZZ\to \ell\ell +\jets$
can be exploited efficiently. In order to do this one has to reduce the 
enormous $W+\jets$ and $Z+\jets$ backgrounds by kinematic cuts.
Henceforth the discussion will be for the $WW$ final state; the $ZZ$ state is similar.
The first step is to reconstruct the $W$ decay to jets.
 Full and fast simulations of the ATLAS detector were used and are in
 good agreement~\cite{AtlasPhysTDR}. At large values of $m_H$ the jets
 from the $W$ decay tend to overlap and several methods were used to
 reconstruct the $W$. In one method,
jets were found using a cone of size $\Delta R=0.2$  and $E_T>50\,\GeV$.
The invariant mass of the di-jet system was then computed by adding
the four-momenta  of the calorimeter cells assuming that each cell is
massless. The di-jet system is required to have $E_T>150\,\GeV$
 This algorithm 
reconstructs $W\to \jets$ with an efficiency of about 60\% and
a $W$ mass resolution of 6.9 GeV for $W's$ produced in the decay of 
$1\,\TeV$ Higgs bosons.   
The mass resolution improves to 5 GeV  at low luminosity where pile up
is unimportant. The dijet system is then required to have a mass
within $2\sigma$ of the nominal $W$ mass. In addition the events are
required to have a lepton with $p_T>50\,\GeV$ and $\Etmiss > 50\,\GeV$.
These cuts applied to
the $W(\to \ell\nu)+\jets$ sample with $p_T(W)>200\,\GeV$ reduce the
rate 
for this process by a factor of 600 and brings it to a
level approximately equal to that from $t\overline{t}$ production; 
$t\overline{t}\to Wb W\overline{b}$.

\begin{table}
\caption[]{$H \to WW \to \ell\nu jj$ signals and backgrounds,
for $m_H=1$~TeV, before and after cuts in the forward region (see text).
The rates are computed for an integrated luminosity of 
$30 \fbi$
and a lepton efficiency of 90\%. Only the $qq \to Hqq$ contribution 
to the signal
is included. Table from  ATLAS simulation.
\label{taggingtable}}
\begin{center}
\begin{tabular}{|l|r|r|r|}
\hline
Process&Central&Jet&Double\\
&cuts&veto&tag\\
\hline\hline
$H\to WW$&222&143&73\\
\hline
$t\overline{t}$&38300&2800&85\\
\hline
$W+jets$&15700&6900&62\\
\hline
\end{tabular}
\end{center}
\end{table}

After these cuts, the backgrounds from $W+\jets$ and 
$t\overline{t}$ are still larger than the signal from $H\to WW$ and topological
cuts are required.  
The process $qq\to Hqq$ produces the Higgs boson in 
association with jets at large rapidity.
These jets can be used as a tag to reject background. 
This forward jet tag will cause some loss of signal since the $gg\to H$
process lacks these forward jets. 
Hence it is only effective for high mass Higgs bosons where the $qq\to Hqq$ process is a significant part
of the cross section. 
Since the Higgs is produced by color singlet $W$ bosons, the central
region in rapidity should have less jet activity in it for Higgs events
than for the background, particularly for that from $t\overline{t}$.  At
low luminosity, requiring that the events have no additional jets (apart
from the ones that make up the W candidate) with $E_T>20$ GeV and
$\abseta< 2$ loses approximately 35\% of the signal and reduces the
background from $W+jets$ ($t\overline{t}$) by a factor of 2.5 (12).

Forward jet tagging was investigated in ATLAS as follows. 
Clusters of energy of size $\Delta R=0.5$ were found in the
region $2<\abseta <5$. Events from the pile up of minimum bias 
events have jets in these regions so the threshold on $E_T$ of
the jet must be set high enough so that these jets do not generate 
tags in the background. If the individual calorimeter cells are
required to have $E_T>3$ GeV, then there is there is a 
4.6\% (0.07\%) probability that the pile up at high luminosity  will
contribute a single (double) tag to an event that would otherwise
 not have one for tagging jets with $E_T>15$ GeV and $E>600$ GeV.
The requirement of a  double tag is then applied to the 
signal from a Higgs boson of mass 1 TeV and the various
backgrounds. The pile up contributions are included and the event rates for a
luminosity of $30$ $\fbi$ shown in table~\ref{taggingtable}.
 The effect of a change in the tagging criteria
can be seen in Fig.~\ref{atlashww} which shows the variation of the
shape in the background. 
The $ZZ$ final state is cleaner as there is no $t\overline{t}$ background 
but the event rates are much smaller. 

A separate study was performed by the CMS group~\cite{cms,cmslnujj}. Here two 
tagging jets with $\abseta >2.4$, $E_T>10$ GeV and $E>400$ GeV are
required. Two central jets are required with in 
invariant mass within 15 GeV of the $W$ or $Z$ mass. For the $ZZ$ case,
the $Z$ is reconstructed from $e$ or $\mu$ pairs with 
invariant mass within 10 GeV of the Z mass; each lepton has
$p_T>50\,\GeV$ and the pair has $p_T>150\,\GeV$. For the $WW$ case, 
at least 150 GeV of missing $E_T$ is needed and the charged lepton
from the $W$ has $p_T>150\,\GeV$.  

It can be seen from  Table~\ref{taggingtable} that it will be 
possible to extract a signal although there are large uncertainties on
the estimated background.  However, other kinematic quantities may be used to
further discriminate between the signal and the background.
The $ZZ$ final state is cleaner as there is no $t\overline{t}$ background 
but the event rates are much smaller. 

\subsubsection{Measurements of Higgs properties}

\begin{figure}
\dofig{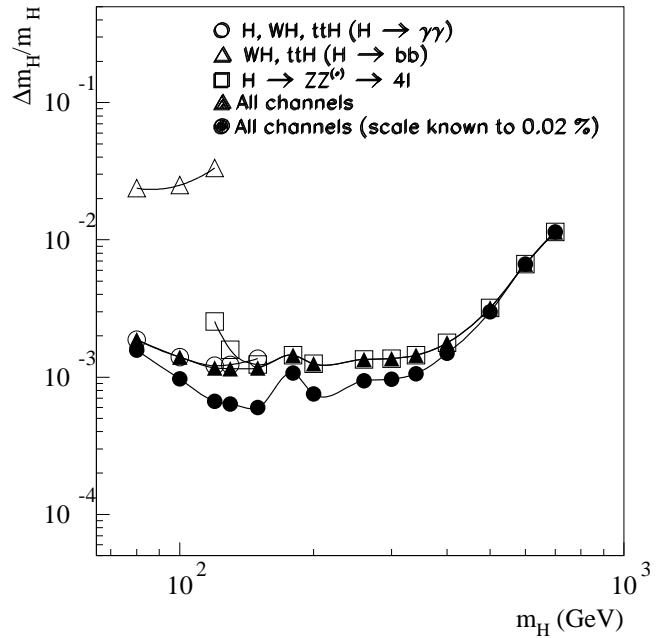}
\caption{Expected ultimate errors on the Higgs mass in ATLAS. 
From Ref.~\cite{AtlasPhysTDR}. \label{hmass}}
\end{figure}

A Standard Model Higgs should have a mass between about $113.5\,\GeV$ and
$212\,\GeV$~\cite{lepfits}. Over this mass range the branching ratios
and other properties of the Higgs vary rapidly, but they are precisely
predicted in terms of the mass. In the $\gamma\gamma$ and four-lepton
channels, the mass resolution is typically 1\%, and the energy scale can
be calibrated to better than  0.1\% using $Z \to ee$ and $Z \to \mu\mu$ events.
Fig.~\ref{hmass} shows that the mass can be measured to $\sim 0.1\%$
for all favored  masses~\cite{AtlasPhysTDR}.

Higgs branching ratios cannot be determined directly at the LHC, but it
is possible to infer combinations of couplings from measured rates.
 The dominant Higgs production mode is $gg \to H$, so
measurements of the inclusive $H \to \gamma\gamma$ and $H \to ZZ^*$
cross sections as discussed above can be used to determine the product
of the $Hgg$ and the $H\gamma\gamma$ or $HZZ$ couplings. The $Hgg$
coupling in turn is related to the $Ht\bar t$ one. 

More information can be obtained by making use of the $WW$ fusion
process. About 10\% of the total cross section in this mass range comes
from $qq \to qqH$ via the exchange of two virtual $W$ bosons:

\dofigb{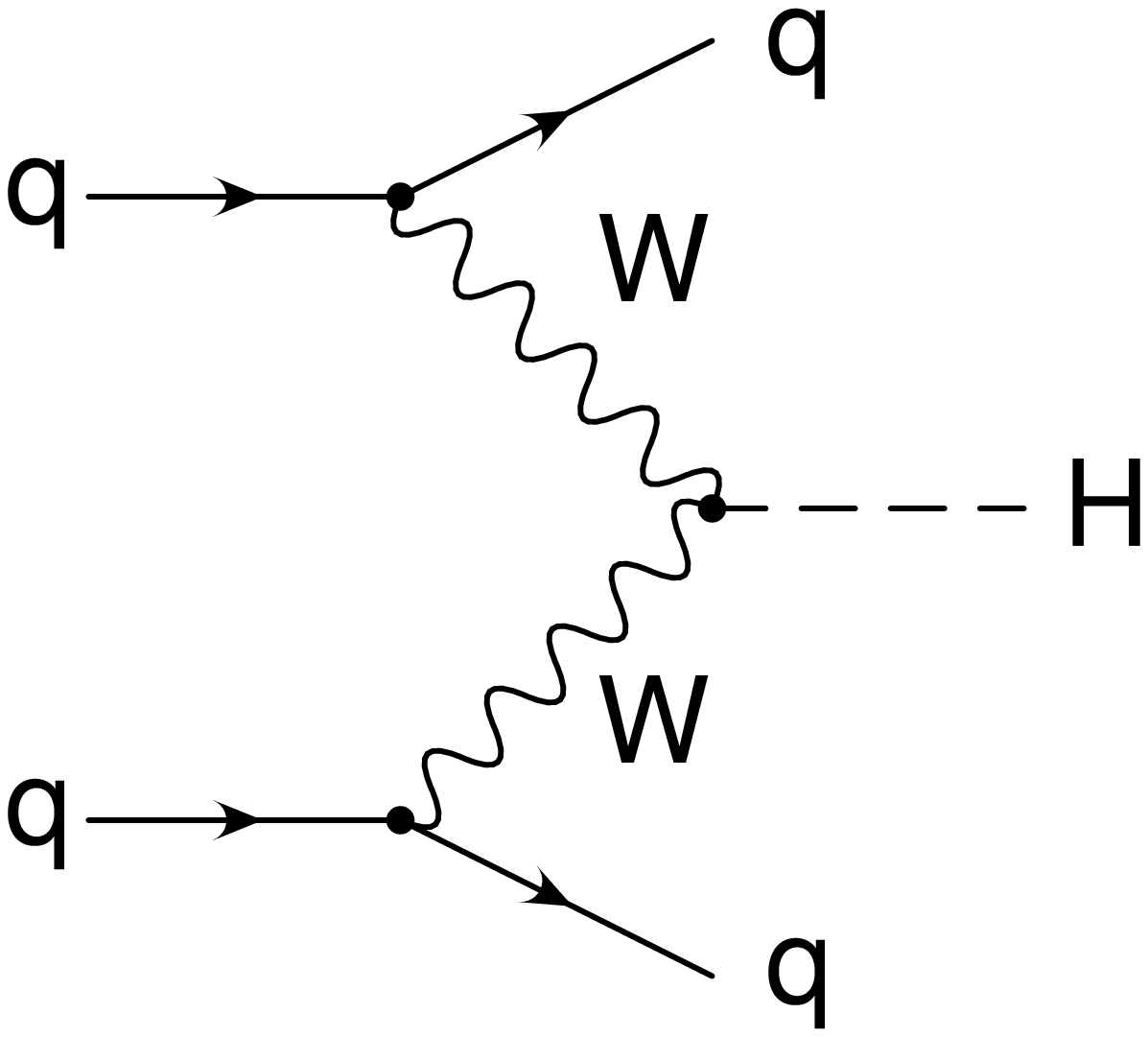}

\noindent The probability that a virtual $W$ is radiated carrying a
fraction $x$ of the momentum of the incoming quark behaves like $dx/x$
at small $x$, so the outgoing quarks typically have large momentum.
Thus, the $WW$ fusion process can be identified by requiring high-energy
jets with $p_T \sim M_W$ in the forward calorimeters and no additional
QCD radiation in the central region. These requirements greatly reduce
the QCD backgrounds. Exploitation of this process requires a detailed
understanding of the forward jet tagging. Complete simulations of
these have not yet been completed.

\begin{figure}
\dofig{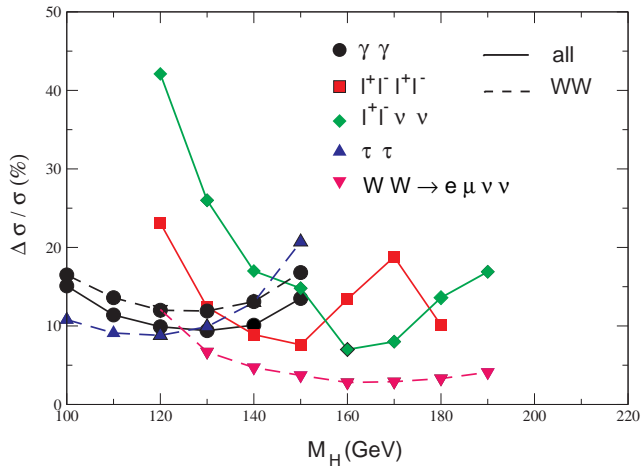}
\caption{Estimated statistical errors on the cross sections for
inclusive Higgs production and production via $WW$ fusion with decays
into various modes for $100\,\fbi$ ($30\,\fbi$ for the
$\ell^+\ell^-\nu\bar\nu$ mode).
Based on Ref.~\cite{Zeppenfeld:2000td}. \label{herr}}
\end{figure}

The estimated statistical errors on the cross sections for a number of
Higgs production and decay channels are shown in Fig.~\ref{herr}. These
have been calculated by applying selection criteria developed for
various Higgs searches separately for ATLAS and CMS, calculating the
errors $\Delta\sigma/\sigma = \sqrt{S+B}/S$, and combining the
results~\cite{Zeppenfeld:2000td}.  The $WW \to H \to \tau\tau$ channel is
reconstructed using the fact that $WW$ fusion provides a transverse
boost to the Higgs, so that one can project the $\Etmiss$ along the two
measured $\tau$ directions and reconstruct the mass, as discussed in
connection with the search for $A \to \tau\tau$ below. Note that for
each Higgs mass there are several channels that can be measured with
statistical errors between 5\% and 20\%. 

It is of course necessary to correct these measurements for acceptance.
For a process like $gg \to H \to \gamma\gamma$ or $gg \to H \to
\ell^+\ell^-\ell^+\ell^-$ this is relatively straightforward. The signal
is a narrow bump on a smooth background, and the losses from geometrical
acceptance, isolation cuts, etc., are relatively small and understood.
The acceptance and background corrections for the forward jet tags
needed to select $WW$ fusion are more difficult to estimate. Ultimately
it will be necessary to vary the cuts and compare the results with both
Monte Carlo event generators and matrix element calculations. The $WW
\to H \to \tau\tau$ channel also has difficult corrections related to
the $\tau$ identification and measurement. After the corrected cross
sections are obtained, they must be compared with perturbative QCD
calculations of the cross sections to determine the relevant combination
of couplings. These calculations are known to NLO in all cases and have
recently been calculated to NNLO for the $gg \to H$ processes.

Studies of this program of measurements are actively underway in both
ATLAS and CMS. Reliable estimates of the expected errors are not yet
available, but it seems plausible that measurements for several channels
will be possible with errors in the 10\%--20\% range. This will provide
a significant amount of information on the couplings of the Higgs.

\subsubsection{Summary of Standard Model Higgs}

The LHC at full luminosity will be able to probe the {\it entire range
of Higgs masses} from the lower limit set by LEP up to the value where
it is no longer sensible to speak of an elementary Higgs boson. The
search mainly relies only on final states that one is confident will be
effective:  $\gamma\gamma$, $4\ell$ and $2\ell \nu\overline{\nu}$.
Additional final states that afford an excellent chance of having a
signal will be exploited to support these: $b\overline{b}$ with an
associated lepton tag at low mass, and $\ell\nu+jets$ and $\ell\ell
+jets$ at high mass.  The failure to find a Higgs boson over this range would
therefore enable the Standard Model to be ruled out.  The Higgs sector
then either consists of non-standard Higgs bosons or the electroweak
symmetry breaking occurs  via some strongly coupled process that will
manifest itself in the study of $WW$ scattering. The next subsection is
devoted to an example of the former type.

\subsection{SUSY Higgs}
As stated above the minimal supersymmetry model (MSSM) has three neutral and 
one charged Higgs bosons;  $h$, $H$, $A$ and $H^\pm$.  
These arise because supersymmetric models, unlike the Standard Model, 
need different
Higgs bosons to generate masses for the up and down type quarks. 
In the Standard Model one parameter,
the Higgs mass, is sufficient to fully fix its properties. In the 
MSSM, two parameters are needed.
These can be taken to be the mass of $A$ 
and the ratio ($\tan \beta$) of the vacuum expectation values of
the Higgs fields that couple to up-type and down-type quarks. 
If $\tan \beta$ is $O(1)$, then coupling of the top quark to Higgs
bosons ($\lambda_t$) is much larger than that of bottom quarks 
($\lambda_b$) as is the case in the Standard Model. 

None of these Higgs bosons has been observed, so we need consider 
only the regions of parameter space not yet excluded.
At tree level the masses of $h$ and $H$ are given in terms of the mass
of $A$ and $\tan \beta$.
The charged Higgs boson $H^\pm$ is heavier than $A$ ($M_{H^{\pm}}^2\sim
M_A^2+M_W^2$).  The $H$ is heavier than the $A$, while the $A$ and $H$
are almost degenerate at large values of $M_A$,
The mass of the lightest boson, $h$,
increases with the mass of $A$ and reaches a plateau for $A$ heavier than
about 200 GeV. The actual values depend on the masses of the other
particles 
in the theory particularly the top quark~\cite{MSSMmasses}.
There is also a dependence (via radiative corrections) on the unknown
masses and other parameters
of the other supersymmetric particles. This dependence
is small if these particles are heavy, so it is conventional to assume
that this is the case.

In the limit of large $A$ mass, the couplings of the Higgs bosons are
easy 
to describe. The couplings of $h$ become like those of
the Standard Model Higgs boson. The couplings of $A$ and
$H$
 to charge 1/3 quarks and leptons are enhanced at large
$\tan \beta$ relative to those of a Standard Model Higgs boson of the
same 
mass. However, $A$ does not couple to gauge boson pairs at lowest order
and the coupling of $H$ to them 
is suppressed at large $\tan \beta$ and large $M_A$.  The decay modes
used 
above in the case of the Standard
Model Higgs boson can also be exploited in the SUSY Higgs case. $h$
can be 
searched for in the final state $\gamma\gamma$, as the
branching ratio approaches that for the Standard Model Higgs in the
large 
$M_A$ (decoupling) limit.

The decay $A\to \gamma\gamma$ can also be exploited. This has the advantage 
that, because $A\to ZZ$ and $A\to WW$ do not occur,
the branching ratio is large enough for the signal to be usable for values of
$M_A$ less than $2m_t$~\cite{snowgamgam}.
The decay $H\to ZZ^*$ can be exploited, but at large values of 
$M_H$ the decay $H\to ZZ$, which provides a very clear signal for the
Standard Model Higgs, is useless owing to its very small branching ratio,
The channel  $t\overline{t}h\to t\overline{t} b\overline{b}$
 can also be exploited.

\begin{figure}[t]
\dofig{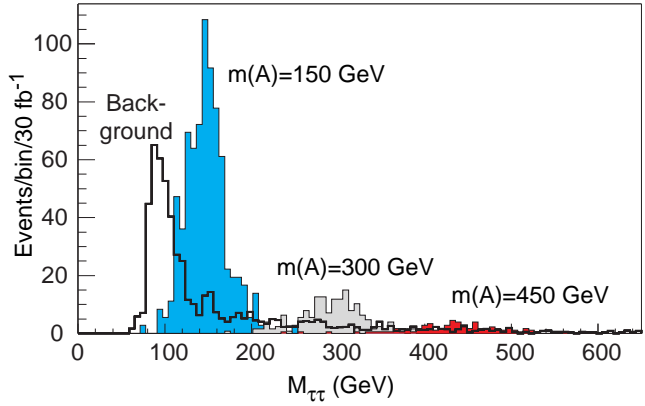}
\caption[]{Reconstructed $\tau\tau$ invariant mass after projecting measured
$\Etmiss$ along the observed $\tau$ directions. ATLAS from  
Ref.~\cite{cavalli99}.
\label{atlastautau}}
\end{figure}

\begin{figure}[t]
\dofig{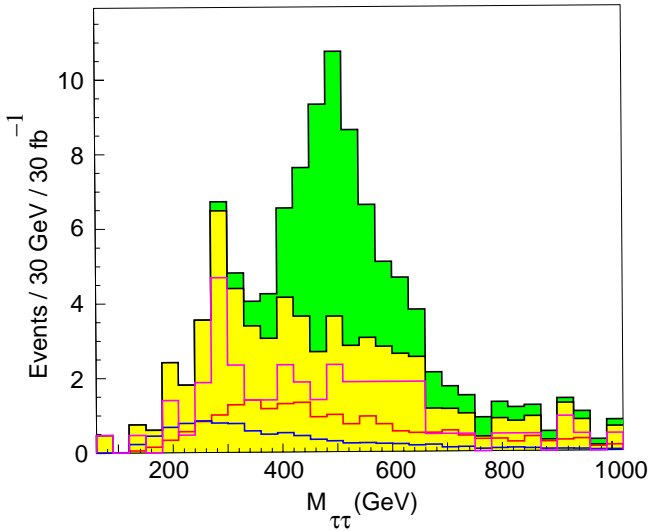}
\caption[]{Reconstructed $\tau\tau$ invariant mass for $M_{H}=500
  \GeV$ and $\tan\beta=25$ from a CMS simulation.
Ref.~\cite{ritva}.
\label{cmstautau}}
\end{figure}

In addition to these decay channels, several other possibilities open
up due 
to the larger number of Higgs bosons and 
possibly enhanced branching ratios. The most important of these 
are the decays of $H$ and $A$ to $\tau^+\tau^-$ and $\mu^+\mu^-$, $H
\to hh$,
 $A \to Zh$ and
$A\to t\overline t$.

It is important to remark that the effect of supersymmetric particles
is ignored in this section. That is, the possible decays of Higgs
bosons to supersymmetric particles are not considered and
supersymmetric particles have been assumed to be heavier than 1 TeV,
so that their effects on branching ratios and production rates 
 via radiative corrections
are ignored. Some effects of these decays have been 
studied~\cite{AtlasPhysTDR}; 
the section below on supersymmetry discusses the
case where Higgs bosons can be produced in the decays of
supersymmetric particles.

\subsubsection{$H/A \to \tau\tau$}

In the MSSM, the $H\to\tau^+\tau^-$ and $A\to\tau^+\tau^-$ rates are strongly
enhanced over the Standard Model if $\tan\beta$ is large, resulting in
the 
possibility of 
observation over a large region of parameter space.  
The $\tau^+\tau^-$ signature can be searched for either in a lepton$+$hadron
final state, or an $e+\mu$ final state.  As there are always neutrinos to
contend with, mass reconstruction is difficult, and $\Etmiss$ resolution is
critical.  In ATLAS, at high luminosity this resolution is 
$$
\sigma(\slashchar{E}_{T,x})=\sigma(\slashchar{E}_{T,y})=0.46\sqrt{\sum E_T}
$$
where all energies are measured in GeV.
Irreducible backgrounds arise from Drell-Yan tau pair
production, $t\overline{t}$  and $b\overline{b}$ 
decays to $\tau\tau$.
Both CMS~\cite{cmstautau} and ATLAS~\cite{atlastautau} have 
studied $\tau^+\tau^-$ final states using full
simulation.

%
%
For the lepton$+$hadron final state, there are additional reducible 
backgrounds from events with one hard lepton plus a jet that is misidentified
as a tau.  In the CMS and ATLAS studies,
events were required to have one isolated lepton with $p_T >15-40$~GeV 
depending on $m_A$ (CMS) or $p_T >24 $~GeV (ATLAS)
within $|\eta|<2.0 (2.4)$ and one tau-jet candidate
within $|\eta|<2.0 (2.5)$. 

ATLAS required that the tau jet have $E_T> 40$~GeV, that the radius of
the jet computed only from the EM cells be less than 0.07; that less
than 10\% of its transverse energy be between $R=0.1$ and $R=0.2$ of its
axis; and again, that exactly one charged track with $p_T>2$~GeV point
to the cluster.  The CMS and ATLAS selections are about 40\%(26\%)
efficient for taus, while accepting only $1/100$ ($1/400$) of ordinary
light quark and gluon jets.

CMS vetoed events having other jets with $E_T > 25$~GeV within $|\eta|<2.4$ (this
reduces the $t\overline{t}$ background); while ATLAS used cuts on
$\Etmiss$, the transverse mass formed from  the lepton and $\Etmiss$, and
the azimuthal angle between the lepton and the tau-jet. 
The mass of the Higgs may be
reconstructed by assuming the neutrino directions to be parallel to those of
the lepton and the tau-jet.  Resolutions of 12 and 14~GeV (Gaussian part) 
are obtained by ATLAS and CMS for $m_A=100$~GeV.  The reconstructed
Higgs peaks  as simulated by ATLAS for several masses are shown in
Fig.~\ref{atlastautau}; a CMS simulation is shown in Fig.~\ref{cmstautau}.

Both ATLAS and CMS find the sensitivity in the $e+\mu$ final state to be
less than for the lepton$+$hadron final state, owing to its smaller 
rate and less favorable decay kinematics.  

Taking the lepton$+$hadron and $e+\mu$ modes together, for the sum of
 $H$ and $A$ decays, both ATLAS and CMS find that the large region 
of parameter space corresponding to
$\tan\beta \simge 6$ at $m_A = 125$~GeV rising to
$\tan\beta \simge 30$ at $m_A = 500$~GeV may be excluded at the 5$\sigma$
confidence level with $30\,{\rm fb}^{-1}$. 
ATLAS also finds some sensitivity to $\tan\beta \simle 2$ for
$125 < m_A <350$~GeV at very high integrated 
luminosities ($300 \,{\rm fb}^{-1}$).  

\subsubsection{$H/A \to \mu\mu$}

The branching ratio for $H$ (or $A$) to $\mu^+\mu^-$ is smaller than that to $\tau^+ \tau^-$ by a factor of
$(m_{\mu} /m_{\tau} )^2$. The better resolution available in this
channel compensates to some extent for this  and the $\mu^+ \mu^-$ mode can be useful for large values of $\tan \beta$.
A signal of less statistical significance than that  in the $\tau^+ \tau^-$ could be used to confirm the discovery and make a 
more precise measurement of the mass and production cross section.
The ATLAS analysis~\cite{AtlasPhysTDR} requires two isolated muons 
with $p_T> 20$ GeV and $\abseta <2.5$. The background from $t\overline{t}$ events
is rejected by requiring $\Etmiss <20 (40)$ GeV at low (high)
luminosity. 
A jet veto could be employed to reduce this background
further, but this is ineffective at reducing the remaining dominant 
background for $\mu^+ \mu^-$ pairs from the Drell-Yan  process.
A cut on the transverse momentum of the muon pair, requiring it to 
be less  than 100 GeV,  reduces the $t\overline{t}$ background further. 
The remaining background 
is very large within $\pm 15 $ GeV of the Z mass. Above this
region the signal appears as a narrow peak in the $\mu^+\mu^-$ mass 
spectrum. In this  region the signal will be
statistically significant if $\tan\beta$ is large enough but it
appears as a 
shoulder on the edge of a steeply falling
distribution which may make it more difficult to extract a signal.

\begin{figure}
\dofig{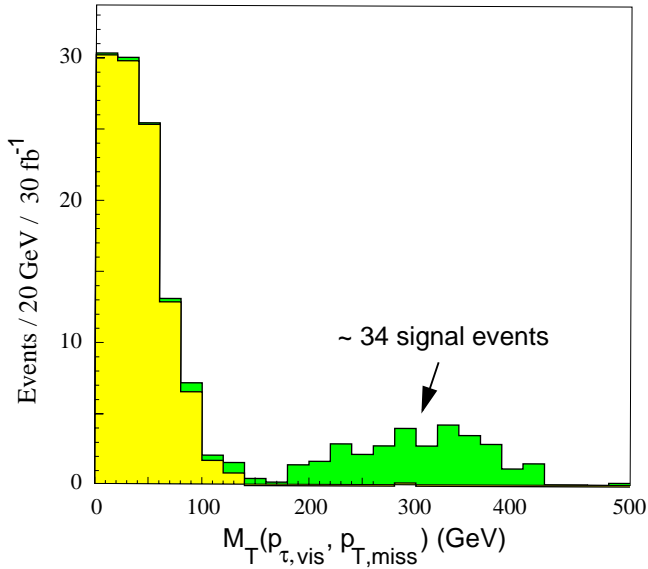}
\caption[]{The transverse mass distribution of signal and backgrounds
for a charged Higgs search using $30 \fbi$. The couplings are
determined in the MSSM with $M_{H^+}=409 \GeV$  and $\tan\beta=40$. From Ref.~\cite{cms2000045}.
\label{chargeh}}
\end{figure}

The significance of the signal in this channel is determined by the $\mu^+ \mu^-$ mass resolution and the intrinsic width of the 
Higgs resonance. The mass resolution in ATLAS is approximately $0.02m_A$ 
and is  $0.013m_A$ in CMS~\cite{cmsmumu}.
At large $\tan \beta$, the masses of $A$ and $H$ are  almost
degenerate 
and they cannot be resolved from each other.  The natural
width of $A$ is proportional to $\tan^2\beta$ and is approximately 3
GeV 
for $\tan\beta=30$ and $M_A=150$ GeV. The mode will
provide a $5\sigma$ signal for a region in the $M_A-\tan\beta$ plane 
covering $M_A>110$ GeV and $\tan\beta>15$ for an integrated
luminosity of $10^{5}$ $\fbi$ .\footnote{The CMS event rates appear
  larger 
than the ATLAS
ones. CMS added the $A$ and $H$ rates whereas the ATLAS numbers
correspond 
to the $A$ alone.}

\subsubsection{$A\to\gamma\gamma$}

Gluon fusion ($gg \to A$) via top and bottom quark 
triangle loop diagrams is the dominant production process if $\tan\beta \simle
4$; while for large $\tan\beta$ ($\simge 7$) $b$-quark fusion dominates.   
For $\tan\beta \approx 1$ and 170 GeV $< m_A < 2m_t$ 
the branching fraction of $A \to \gamma\gamma$ is between
$5 \times 10^{-4}$ and $2 \times 10^{-3}$.
The backgrounds considered are QCD photon production, both the irreducible
two-photon backgrounds ($q \overline{q} \to\gamma \gamma $ and 
$ g g \to\gamma \gamma $) and 
the reducible backgrounds with one real photon
($q \overline{q} \to g \gamma$, $q g \to q \gamma$,  
and $g g \to g \gamma$). In the ATLAS study~\cite{AtlasPhysTDR},
both photons were required to have  $|\eta| < 2.5$, one with  $P_T>125$
GeV and the other with   $p_T>25$ GeV.
Both photons are required to be isolated. The signal is effective at
small values of $\tan\beta$ for $2m_t<M_A<200$ GeV. 

\subsubsection{Search for Charged Higgs}

In extensions of the Standard Model with charged Higgs bosons $H^\pm$, such as
in the MSSM, the decay $t \to b H^\pm$ may compete with the standard 
$t \to b W^\pm$ if kinematically allowed.  The $H^\pm$ decays to $\tau\nu$ or
$c \overline s$ depending on the value of $\tan\beta$.  Over 
most of the range $1 < \tan\beta < 50$, the decay mode
$H^\pm \to \tau\nu$ dominates.  The signal for $H^\pm$ production is thus an
excess of taus produced in $t\overline t$ events.

Both ATLAS~\cite{atlascharg} and CMS~\cite{cmscharg} have 
investigated the sensitivity to this excess.  
Top events with at least one isolated high-$p_T$ lepton are selected, and
the number having an additional tau compared with the number having an
additional $e$ or $\mu$.  Both studies used $b$-tagging to reduce the
backgrounds to top production.
Taus were identified
in a way very similar to that described earlier (in the section on
$A,H \to \tau \tau$
searches). 
The uncertainty in the tau excess is estimated to be $\pm 3$\%, dominated
by systematics. 
For an integrated luminosity of $10\,\fbi$, both ATLAS and CMS conclude that
over most of the $\tan\beta$ range, a signal can be observed at the $5\sigma$
level for $m_{H^\pm} < 130$~GeV, which corresponds to the region
$m_A \simle 120$~GeV in the $m_A,\tan\beta$ plane. 

If a charged Higgs boson has larger mass than the top then it cannot
be produced in the decay of a top quark. In this case the relevant
production mechanism is the $gb \to H^- t$~\cite{cms2000045,coadou}. 
The signal can be searched for via the decay $H^+ \to
\tau \nu$. The tau is searched for via its hadronic decay which gives
rise to isolated single hadrons (either $\pi$ or $K$). This track is
required to have $p_T>100$ GeV. Events are then 
required to have a single tagged $b$-jet and two other jets whose masses are
consistent with the decay $t\to Wb \to qqb$ and  $\Etmiss > 100 $ GeV.
 Events with two tagged $b$-jets are vetoed. A transverse mass 
is then formed between the
reconstructed single hadron and the $\Etmiss$. The distribution of
this transverse mass is shown in Fig.~\ref{chargeh} for 
a charged Higgs mass of 500 GeV. The Standard Model background is small. Note
that the peak is below the mass of the charged Higgs. This is due to
the partial cancellation of missing $E_T$ from the two neutrinos in
the decay chain $H^+ \to
\tau \nu \to \pi\nu\nu$.

\begin{figure}
{\centerline{\includegraphics[width=7.0cm,angle=270]{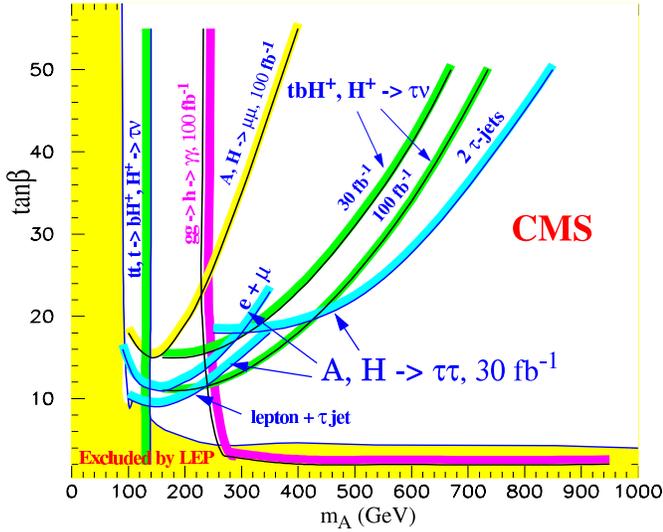}}}
\caption[]{$5\sigma$ discovery  contours for the various processes
used to search for Higgs bosons in the MSSM. This plot assumes no stop
mixing, maximizing the reach of LEP. From Ref.~\cite{denegrimssm}.
\label{cmsfroid}}
\end{figure}

\subsubsection{Other possible Higgs signatures}

Observation of the channel $H \to hh$ would be particularly interesting
as information about two different Higgs bosons and their coupling could
be obtained.  The dominant decay here is  to the final state
$b\overline{b}b\overline{b}$.  However it is not clear how this mode
could be triggered efficiently. If a trigger could be constructed ---
perhaps using soft muons in jets --- then the process is sensitive for
$\tan\beta<3$ and $250<M_A<2m_t$. The channel $H\to hh\to b\overline{b}
\tau^+\tau^-$ can be triggered and is being studied.
 The LEP Higgs limits exclude most of
the accessible region in the MSSM, but these channels might be
observable in more general models.

The decay channel $H\to hh \to \gamma\gamma b\overline{b}$ is
triggerable and was studied~\cite{AtlasPhysTDR}. Events were required to
have a pair of isolated photons with $\abseta<2.5$ and $p_T>20$ GeV and
two jets with $p_T> 15 (30)$ GeV and $\abseta<2.5$ at low (high)
luminosity. One of the jets was required to be tagged as a $b$-jet. No
other jets with $P_T>30$ GeV were allowed in the region $\abseta<2.5$.
The dominant background arises from $\gamma\gamma$ production in
association with light quark jets and is approximately 10 times larger
than the $\gamma\gamma b\overline{b}$ background.  Event rates are very
low, for $M_H\sim 250$ GeV and $m_h=100$ GeV there are about 15 signal
events for 200 $\fbi$ of integrated luminosity.  However the very small
background ($\sim 2$ events for $200\,\fbi$) and the sharp peak in the
$\gamma\gamma$ mass distribution should provide convincing evidence of a
signal.

For large masses, the $A$ and $H$ decay almost exclusively to
$t\overline{t}$. The background in this channel arises from QCD
$t\overline{t}$ production. While this background is very large, a statistically
significant signal can be extracted provided that the background can be
calibrated~\cite{AtlasPhysTDR}. The signal is searched for in the final
state $WWb\overline{b}$ where one W decays leptonically.  For an
integrated luminosity of $30\fbi$ there are about 2000 events for
$M_A\sim 400$ GeV after cuts requiring an isolated lepton (which
provides the trigger) and a pair of tagged $b$-quark jets. The
$t\overline{t}$ mass resolution is of order 15 GeV resulting in
approximately 40000 background events.  The rate for $t\overline{t}$
production is well predicted by perturbative QCD, so it may well be
possible to  establish an event excess but extraction of a
mass for $A$ will be very difficult as there is no observable mass
peak. 
The mode is most likely to be
useful as confirmation of a signal seen elsewhere.

The decay $A \to Zh$ offers another channel where two Higgs bosons might
be observed simultaneously.  The leptonic decay of the $Z$ can be used
as a trigger.  The CMS study requires a pair of electrons (muons) with
$p_T >20 $~(5)~GeV which have an invariant mass within 6 GeV of the $Z$ mass
and a pair of jets with $p_T>40$ GeV. One or two $b$-tags are required
with an assumed efficiency of 40\% and a rejection of 50 against light
quark jets.  The background is dominated by $t\overline{t}$ events.  The
signal to background ratio is quite good for moderate $M_A$ and small
$\tan\beta$, but this region is excluded in the MSSM by the LEP Higgs
limits.

\begin{figure*}[t]
\dofigc{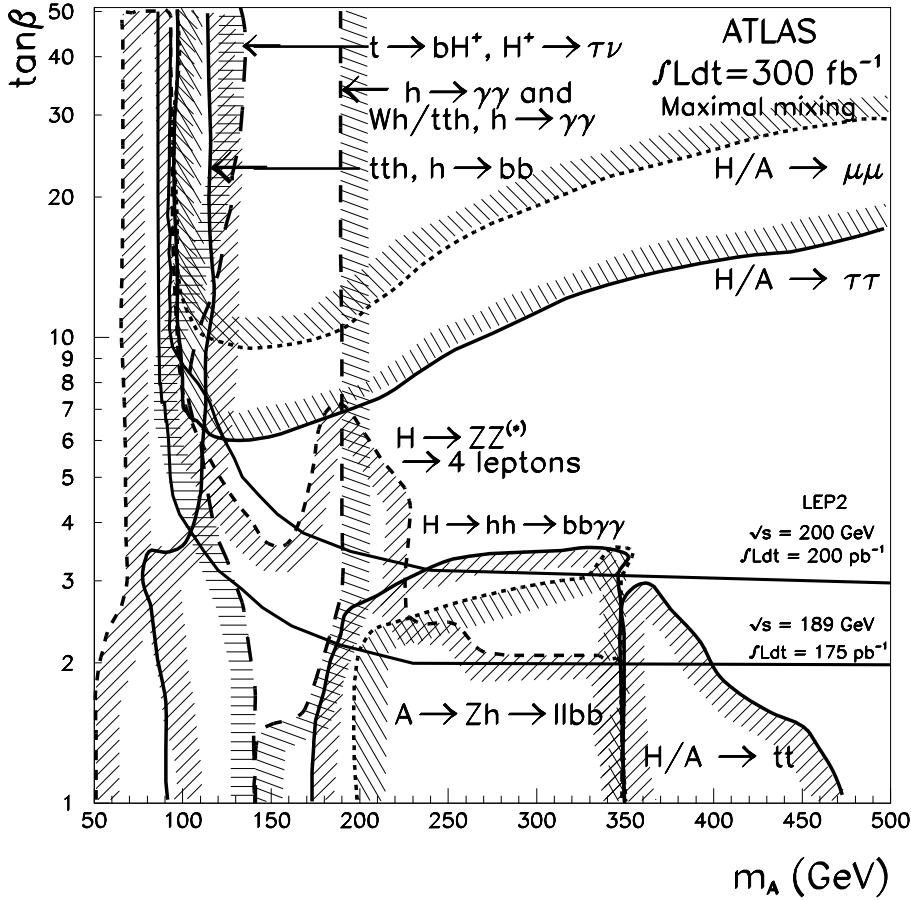}
\caption[]{$5\sigma$ discovery  contours for the various processes
used to search for Higgs bosons in the MSSM. This plot assumes maximal stop
mixing, minimizing the reach of LEP. From Ref.~\cite{AtlasPhysTDR}.
\label{froidfig}}
\end{figure*} 

The positive conclusion of this study is confirmed in
~\cite{AtlasPhysTDR} 
where several values of $M_A$ and $m_h$ were simulated and
it was concluded that a $5\sigma$ signal is observable for an
integrated 
luminosity of $30$ $\fbi$ for $\tan\beta
<2$ and $150 <M_A< 350$. This study included the background
 from $Zb\overline{b}$ events which dominate over the $t\overline{t}$
background at smaller values of $m_A$.

\subsubsection{Summary of Supersymmetric Higgs}

One is confident that the following modes will be effective in 
searching for the MSSM Higgs bosons: $A/H \to \tau^+\tau^-$,
$A/H\to \mu^+\mu^-$, $H^+\to \tau\nu$, $H\to Z Z^* \to 4\ell$, $h\to \gamma\gamma$, 
$A\to Zh\to \ell\ell b \overline{b}$, $H\to h h \to
b\overline{b}\gamma\gamma$  and  $t \to b H^+(\to \tau \nu)$
(discussed in the section on the top quark). In addition, the modes 
$A/H\to t\overline{t}$ and  $h\to b\overline{b}$ 
produced in association with a $W$ or  $t\overline{t}$ may 
provide valuable information. 
The former set of modes are sufficient for either experiment 
to {\it exclude} the
entire $\tan\beta - M_A$ plane at 95\% confidence with $100$ $\fbi$.

Ensuring a $5\sigma$ discovery over the entire $\tan\beta - M_A$ plane 
requires more luminosity. Figs.~\ref{cmsfroid} and \ref{froidfig}
show  what 
can be achieved. 
 The entire plane is 
covered using the modes where one has great
confidence. Over a significant fraction of the parameter space at least two 
distinct modes will be visible.  Over a significant fraction of
the phase space beyond the LEP limit, 
$h\to \gamma\gamma$, $H^+\to \tau \nu_{\tau}$ and $H/A\to \tau\tau$ 
($H/A \to \mu\mu$) 
will be measured.  
The decay of other supersymmetric particles will provide additional
sources
 of $h$.  
Over a significant fraction of SUSY parameter space, 
there is a substantial branching fraction for 
sparticles to decay to $h$. The rate is then such that 
decay $h\to b\overline{b}$ becomes clearly observable above background
and this channel is the one where $h$ is observed first at LHC (see below).

\section{Supersymmetry} 

If SUSY is relevant to electroweak symmetry breaking, then the arguments
summarized above suggest that the gluino and squark masses are less than
${\cal O}(1\,\TeV)$, although squarks might be heavier. As many
supersymmetric particles can be produced simultaneously at the LHC,
a model that has a consistent set of
masses and branching ratios must be used for simulation.
Analysis of the simulated events is
performed without reference to the underlying model.
The SUGRA model~\cite{SUGRA} assumes that
gravity is responsible for the mediation of supersymmetry breaking and
provides a natural candidate for cold dark matter. The GMSB model
~\cite{GMSB} assumes that Standard Model gauge interactions are
responsible for the mediation and explains why flavor changing neutral
current effects are small. Anomaly mediation is always present
~\cite{AMSB}; the AMSB model assumes that it is dominant.

\begin{figure}[t]
\dofig{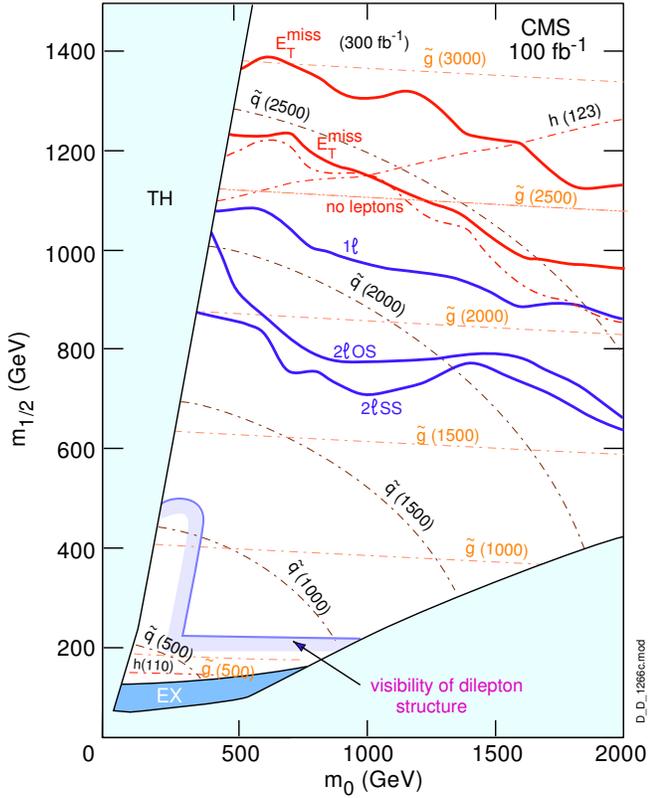}
\caption[]{Plot of $5\sigma$ reach in minimal SUGRA model  for
  $\tan\beta=35$ and $\mu=+$ at LHC with $100\,\fbi$
for $\Etmiss$ inclusive,  $\Etmiss$ with no leptons, $\Etmiss$
 plus one lepton ($1\ell$), opposite sign ($2\ell$OS)
and same-sign ($2\ell$SS) dileptons, and multi-leptons ($3\ell$,$4\ell$).
The region where  a dilepton edge is visible is indicated. From Ref.~\protect\cite{Abdullin:1998pm}.
\label{cms35}}
\end{figure}

\begin{figure}[t]
\dofig{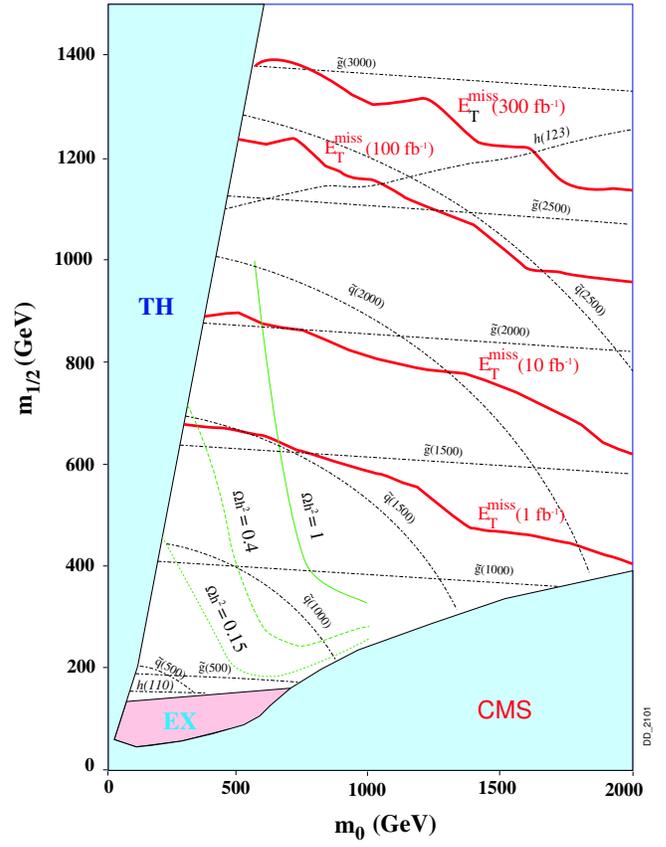}
\caption{Plot of $5\sigma$ reach in minimal SUGRA model  for
  $\tan\beta=35$ and $\mu=+$ at LHC for the  $\Etmiss$ signal for
  various integrated luminosities.
\label{cmslumin}}
\end{figure}

\begin{figure}[t]
\dofig{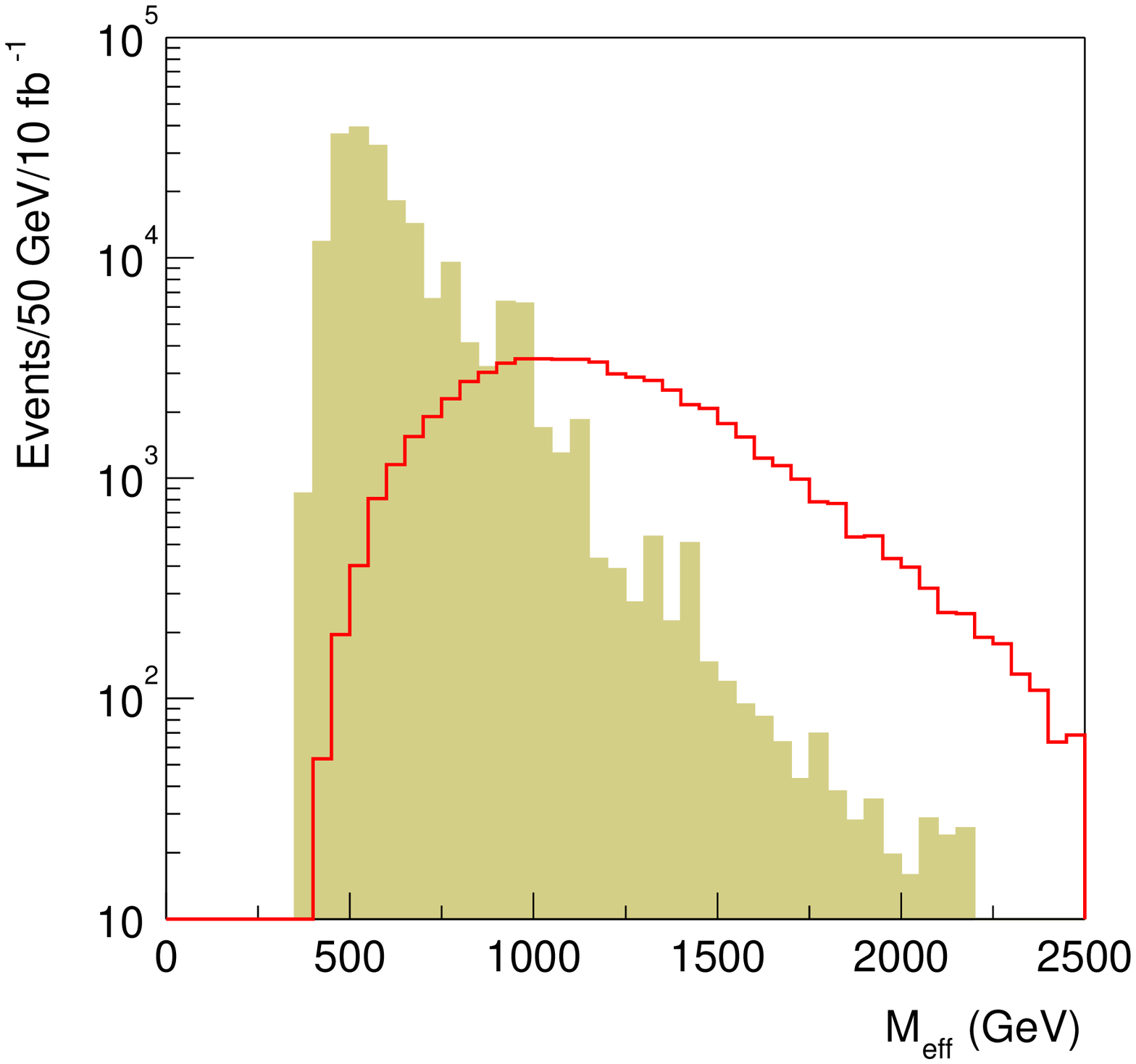}
\caption{$\Meff$ distribution for a SUGRA point with gluino and squark
masses of about $700\,\GeV$ (histogram) and Standard Model background
(shaded) after cuts. Based on Ref.~\protect\cite{Hinchliffe:2000np}.
\label{c10_meff}}
\end{figure}

Gluinos and squarks usually
dominate the LHC SUSY production cross section, which is of order
$10\,\pb$ for masses around 1 TeV.
Since these are strongly produced, it is easy to separate SUSY from Standard
Model backgrounds provided only that the SUSY decays are distinctive. In the
minimal SUGRA model these decays produce $\Etmiss$ from the missing $\lsp$'s
plus multiple jets and varying numbers of leptons from the intermediate
gauginos. Fig.~\ref{cms35} shows the $5\sigma$ reach in
this model at the LHC for  $100\,\fbi$~~\cite{Abdullin:1998pm}
The reach is not very sensitive to the fixed parameters ($A$ and
$\tan\beta$). It is considerably
more than the expected mass range even for $10\,\fbi$ as can be seen
from  Fig.~\ref{cmslumin} which shows how the accessible mass range
depends upon integrated luminosity. This plot also shows the parameter
range over which the model provides a suitable dark matter candidate.

A typical example of the signatures whose reach is shown in
Figure~\ref{cms35},  is the distribution of the
``effective mass''
$$
\Meff = \Etmiss + \sum_{i=1}^4 p_{T,i}
$$
computed from the missing energy and the four hardest jets.  This is
shown in Fig.~\ref{c10_meff} after multijet and $\Etmiss$ cuts for a
SUGRA point~\cite{Hinchliffe:2000np} with gluino and squark masses 
of about $700\,\GeV$.

While the reach in Fig.~\ref{cms35} has been calculated
for a specific SUSY model, the multiple jet plus $\Etmiss$ signature is
generic in most $R$ parity conserving models.  GMSB models can give additional
photons or leptons or long-lived sleptons with high $p_T$ but $\beta<1$,
making the search easier. $R$-parity violating models with leptonic $\neu1$
decays also give extra leptons and very likely violate $e$-$\mu$ universality.
$R$-parity violating models with $\neu1 \to qqq$ give signals at the LHC with
very large jet multiplicity, for which the Standard Model background is not
well known.  For such models, it may be necessary to rely on leptons produced
in the cascade decay of the gluinos and squarks. In all cases, SUSY can be
discovered at the LHC if the masses are in the expected range, and simple
kinematic distributions can be used to estimate the approximate mass
scale~\cite{AtlasPhysTDR}.

\begin{figure}[t]
\dofig{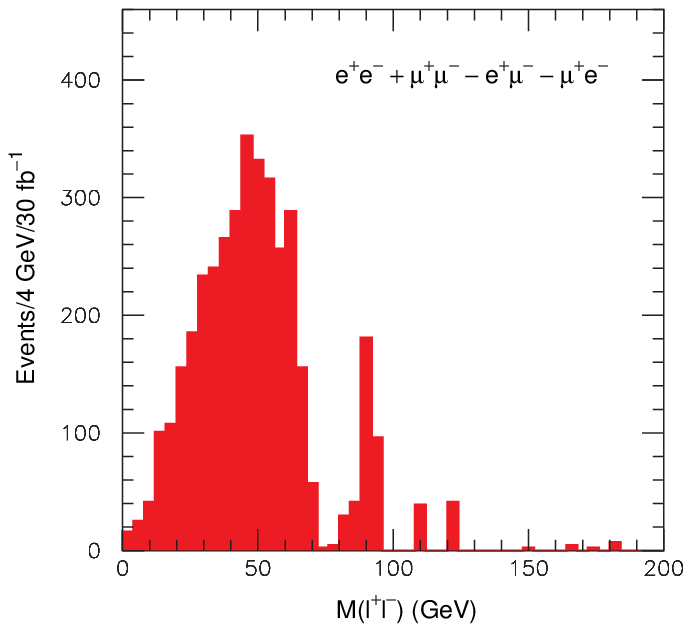}
\caption{Plot of $e^+e^- + \mu^+\mu^- - e^\pm\mu^\mp$ mass distribution for
LHC SUGRA Point~4 with direct $\neu2 \to \neu1\ell\ell$ decay in ATLAS. The
$Z\to\ell^+\ell^-$ signal comes from decays of the heavier gauginos. From
Ref.~\protect\cite{AtlasPhysTDR}.\label{p4_mll}}
\end{figure}

\begin{figure}[t]
\dofig{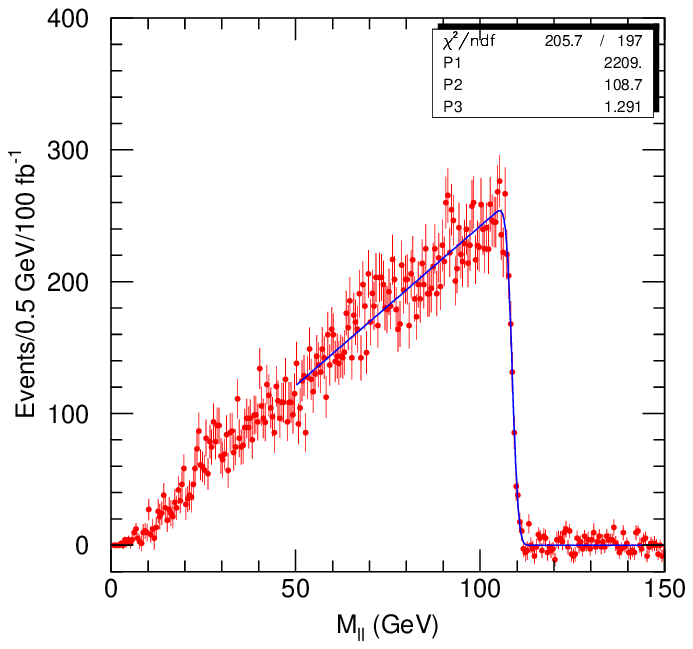}
\caption{Plot of $e^+e^- + \mu^+\mu^- - e^\pm\mu^\mp$ mass distribution for
for LHC SUGRA Point~5 with $\neu2 \to \s\ell^\pm\ell^\mp \to 
\neu1\ell^+\ell^-$ in ATLAS. From Ref.~\protect\cite{AtlasPhysTDR}.\label{c5_mll}}
\end{figure}

\begin{figure}[t]
\dofig{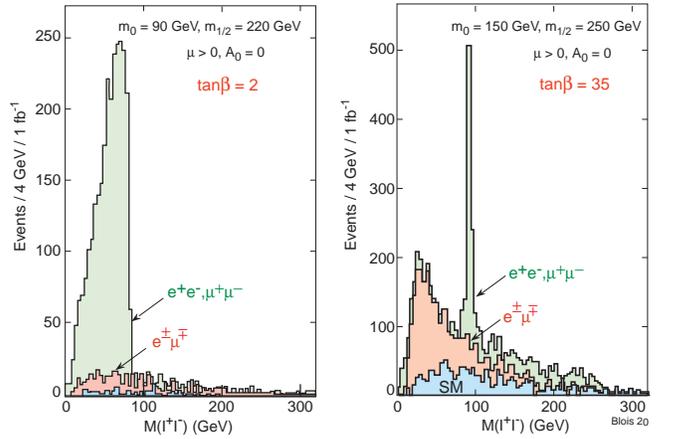}
\caption{Plot of $e^+e^-$, $\mu^+\mu^-$  and  $e^\pm\mu^\mp$ mass
distribution for  SUGRA showing the signal at two points with CMS.
From Ref.~\protect\cite{bloisref}.\label{blois}}
\end{figure}
   
\begin{figure}[t]
\dofig{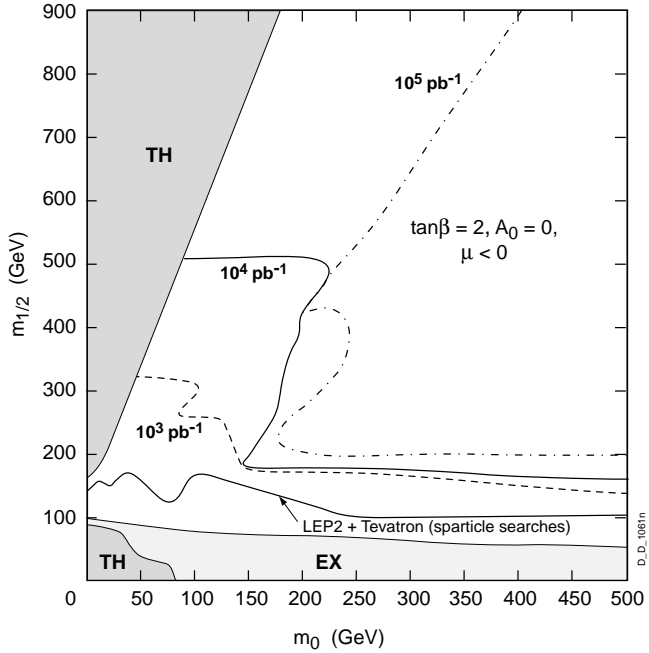}
\caption{Reach for observing dilepton endpoints in SUGRA models with
$1\,\fbi$, $10\,\fbi$ and $100\,\fbi$. Theory (TH) and experimental
constraints are also indicated. From Ref.~\cite{Abdullin:1998pm}. 
\label{cmsll}}
\end{figure}

\subsection{SUGRA Measurements}

\begin{figure}[t]
\dofig{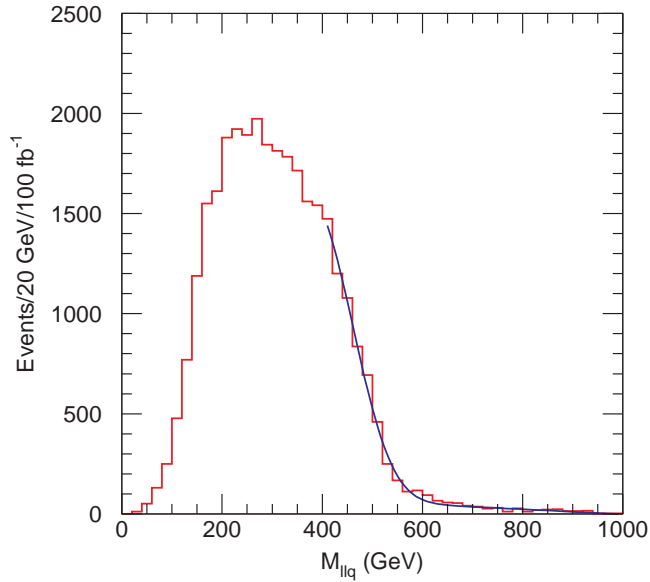}
\caption{Plot of minimum $M(\ell\ell q)$ mass formed from $e^+e^- +
\mu^+\mu^- - e^\pm\mu^\mp$ plus one of two hardest jets at LHC SUGRA
Point~5. The smooth curve shows a
fit used to estimate the error on the endpoint. 
From Ref.~\protect\cite{AtlasPhysTDR}. \label{c5_mllq2}}
\end{figure}

\begin{figure}[t]
\dofig{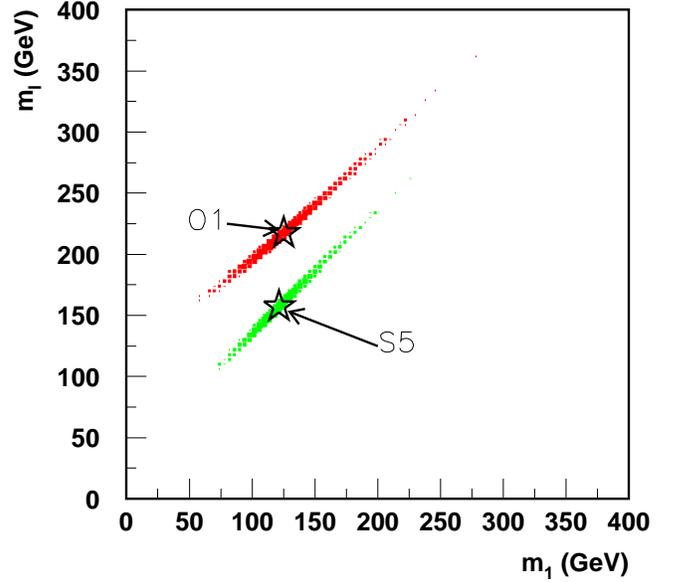}
\caption{Scatter plot of reconstructed values of $m_\ell \equiv M_{\s\ell_R}$
vs.{} $m_1 \equiv M_{\neu1}$ for LHC Point~5 (S5) and for an ``optimized
string model'' (O1) using multiple measurements from the decay chain $\s{q}_L
\to \neu2 q \to\s\ell_R^\pm \ell^\mp q \to \neu1\ell^+\ell^-q$. The stars mark
the input values. From Ref.~\protect\cite{Allanach:2000kt}.\label{camb15}}
\end{figure}

\begin{figure}[t]
\dofig{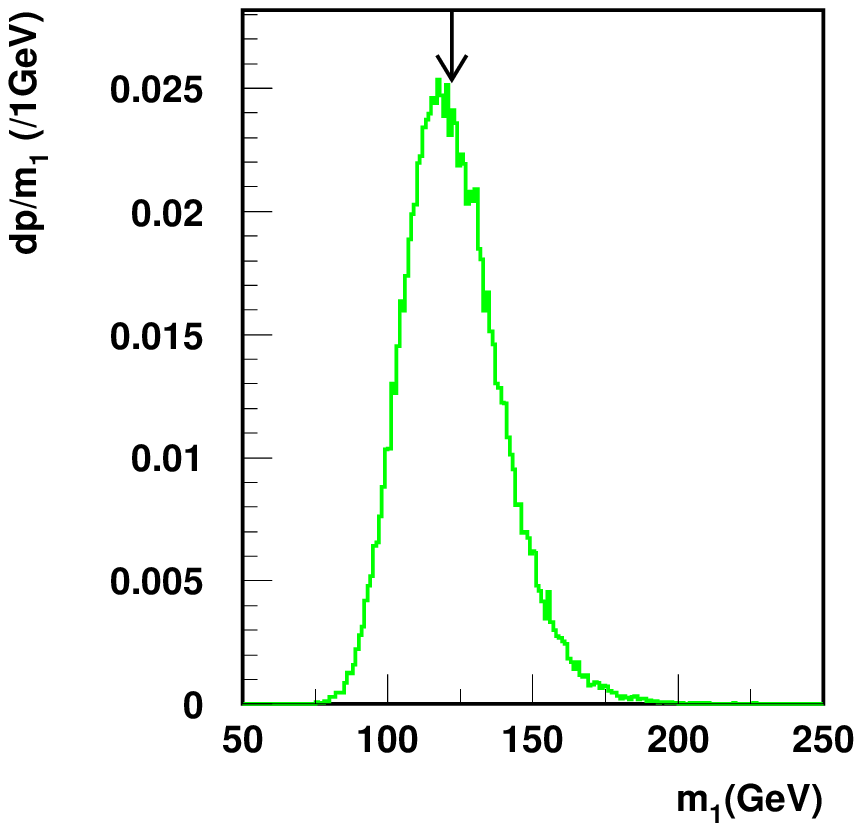}
\caption{Projection of $M_{\neu1}$ in Fig.~\ref{camb15} for LHC Point~5.
From Ref.~\protect\cite{Allanach:2000kt}.\label{camb11}}
\end{figure}

The main problem at the LHC is not to observe a SUSY signal that deviates from
the Standard Model but to separate the many different channels produced by all
the SUSY cascade decays from the produced squarks and gluinos. In SUGRA and
many other models, the decay products of SUSY particles always contain an
invisible $\lsp$, so no masses can be reconstructed directly. One promising
approach is to try to identify particular decay chains and to measure
kinematic endpoints for combinations of visible particles in
these~\cite{Hinchliffe:1997iu}. For example, the $\ell^+\ell^-$ mass
distribution from $\neu2 \to \neu1\ell^+\ell^-$ has an endpoint that measures
$M_{\neu2}-M_{\neu1}$, while the distribution from the two-body decay
$\neu2 \to \s\ell^\pm \ell^\mp \to \neu1 \ell^+\ell^-$ has a different shape 
with a sharp edge at the endpoint
$$
\sqrt{ (M_{\neu2}^2 - M_{\s\ell}^2)(M_{\s\ell}^2 - M_{\neu1}^2) 
\over M_{\s\ell}^2}
$$
 Dilepton mass
distributions~\cite{AtlasPhysTDR} after cuts for an example of each
decay are shown for ATLAS in Figs.~\ref{p4_mll}, \ref{c5_mll} and for CMS in
Fig.~\ref{blois}. The position of the end point is 108.6 GeV in 
Fig~\ref{c5_mll}. 
The flavor-subtraction combination $e^+e^- + \mu^+\mu^- - e^\pm\mu^\mp$
removes backgrounds from two independent decays.
The last plot shows that the signal structure depends
strongly on the choice of parameters. Note that at the small values of
$m_0$ and $m_{1/2}$ shown, the event rates are very large. Such endpoints can
be observed over a wide range of parameters as indicated in
Fig.~\ref{cmsll}~\cite{Abdullin:1998pm}.

When a longer decay chain can be identified, more combinations of masses
can be measured. Consider, for example, the decay chain
$$
\s{q}_L \to \neu2 q \to \s\ell_R^\pm \ell^\mp q \to \neu1\ell^+\ell^-q\,.
$$
For this decay chain, kinematics gives $\ell^+\ell^-$, $\ell^+\ell^- q$, and two
$\ell q$ endpoints as functions of the masses. If a lower limit is imposed on
the $\ell^+\ell^-$ mass, there is also a minimum $\ell^+\ell^- q$ mass. With
suitable cuts all of these can be measured~\cite{AtlasPhysTDR,Bachacou:2000zb}
for the cases considered. An example is the minimum $\ell\ell q$ mass formed
from the dilepton pair shown in Fig.~\ref{c5_mllq2} and one of the two hardest
jets. Since the hardest jets are mainly from squark decays, this smaller mass
should have an endpoint given by the above decay chain at
$$ \sqrt{{\left(M_{\tq_L}^2-M_{\tchi_2^0}^2\right)
\left(M_{\tchi_2^0}^2-M_{\lsp}^2\right) \over M_{\tchi_2^0}^2}} 
$$
In the case shown this endpoint is at 552.4 GeV. The statistical errors on the measured endpoints are typically comparable to
the systematic limits, ${\cal O}(0.1\%)$ for leptons and ${\cal O}(1\%)$ for
jets.  

The set of measurements just described can be used to determine all the masses
in the relevant decay chain. This is most easily done by generating the four
masses at random and comparing the predicted results with the measurements.
Fig.~\ref{camb15} shows a scatter plot of the resulting $\s\ell_R$ and $\neu1$
masses for LHC SUGRA Point~5 and for a similar point in another SUSY model
with this decay chain~\cite{Allanach:2000kt}.  The relations between masses
are determined with good precision, so these two models are easily
distinguished, as can be seen in Fig.~\ref{camb15}. Although the LSP is
invisible, its mass, Fig.~\ref{camb11}, can be measured to ${\cal O}(10\%)$ 
through its effects on the decay kinematics.

\begin{figure*}[t]
\dofige{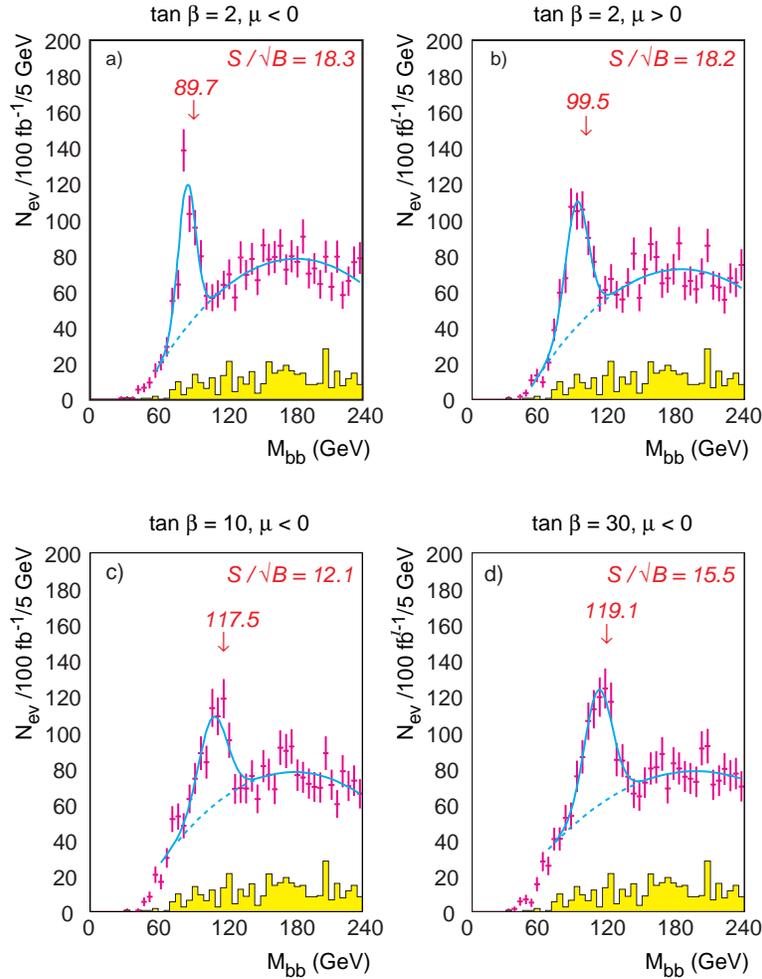}
\caption{Plot of $b \bar b$ dijet mass distribution (points) with $h \to b
\bar b$ signal (solid), SUSY background (dashed), and Standard Model
background (shaded) for various $\tan\beta$. From
Ref.~\protect\cite{cmssusyb} \protect\cite{Abdullin:1998pm} . 
\label{cmshbb}}
\end{figure*}

If the two-body decay $\neu2 \to \lsp h$ is open, it will typically have a
substantial branching ratio; it can be dominant if the $\neu2$ and $\lsp$ are
mainly gaugino and the slepton channel is closed. If events are selected with
multiple jets, large $\Etmiss$, and two tagged $b$ jets, then the decay $h \to
b \bar b$ can be reconstructed. Examples for several points with different
values of $\tan\beta$ are shown in Fig.~\ref{cmshbb}~\cite{cmssusyb}\cite{Abdullin:1998pm}.
Like the dilepton signal, this one can also be combined with additional jets
to provide further information.

\begin{figure}[t]
\dofig{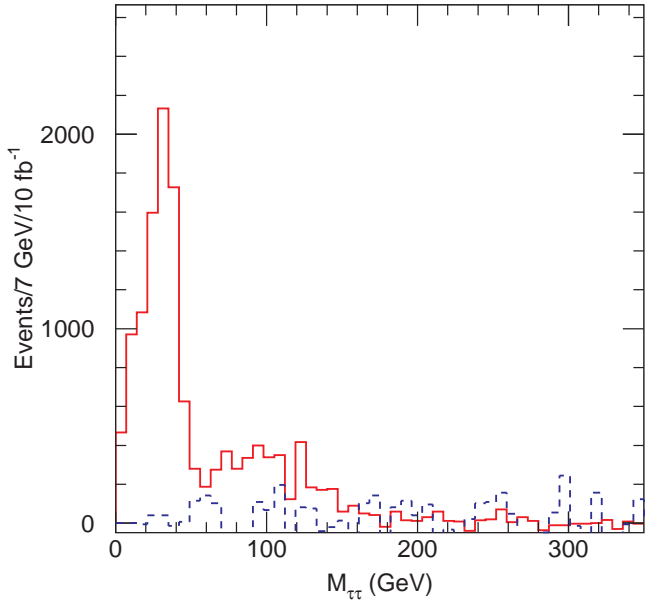}
\caption{Plot of $\tau^+\tau^- - \tau^\pm\tau^\pm$ visible mass distribution
with hadronic $\tau$ decays at LHC SUGRA Point~6. From
Ref.~\protect\cite{AtlasPhysTDR}.\label{p6subtracted}}
\end{figure}

It is also possible that the only two-body decays are $\neu2 \to \ttau_1\tau
\to \lsp\tau\tau$. This can occur naturally in SUGRA if $\neu2 \to \lsp Z$
$\lsp h$, and $\tell\ell$ are all closed but $\tan\beta$ is large enough that
$\neu2 \to \ttau_1\tau$ is open. One analysis of a sample point, LHC SUGRA
Point~6, has been done~\cite{AtlasPhysTDR} using hadronic $\tau$ decays to
determine the $\tau\tau$ mass distribution. Since simple kinematic cuts select
a rather pure SUSY sample with ${\cal O}(1)$ hadronic $\tau$ per event, 
the $\tau$
selection criteria were chosen not to optimize the QCD jet rejection but
rather to select multi-pion decays and so to improve the $\tau\tau$ mass
resolution. The combination $\tau^+\tau^- - \tau^\pm\tau^\pm$ removes most of
the background from misidentified jets. The resulting visible mass
distribution is shown in Fig.~\ref{p6subtracted}. If $\tau$'s could be
measured perfectly, this distribution would have a shape like
Fig.~\ref{c5_mll} with a sharp endpoint at $59.6\,\GeV$. Although the
endpoint is shifted and broadened by the missing neutrinos, measurements at
the $\sim5\%$ level seem possible even in this difficult case. (This
point and similar ones would give a very large contribution to $g_\mu-2$ in
contradiction to Ref.~\protect\cite{Brown:2001mg}.)

Kinematic endpoints are of course only a small part of the data that will be
available from the LHC if SUSY is discovered. One will be able to measure
cross sections, relative branching ratios, and many other kinematic
distributions. For example, in the decay chain $\neu2 \to \tell_R^\pm\ell^\mp
\to \neu1\ell^+\ell^-$, the ratio $p_{T,2}/p_{T,1}$ of the two leptons
contains  information that is independent of the endpoint: 
one lepton will be soft if the slepton is
nearly degenerate with either the $\neu2$ or the $\lsp$. 

\subsection{GMSB Measurements}

In GMSB models the gravitino $\tG$ is very light; the phenomenology is
determined by the nature of the next lightest SUSY particle (NLSP), either the
$\lsp$ or a slepton, and by its lifetime to decay into a $\tG$. GMSB models
generally provide additional experimental handles and so are easier to analyze
than SUGRA models.

\begin{figure}
\dofig{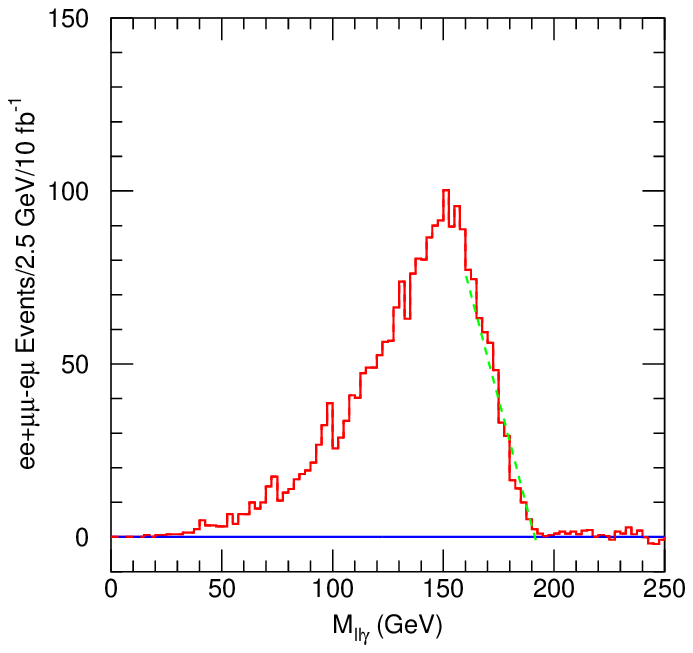}
\caption{$\ell^+\ell^-\gamma$ mass distribution for a GMSB point with a prompt
$\tchi_2^0 \to \tell_R^\pm\ell^\mp \to \lsp\ell^+\ell^- \to \tG\gamma
\ell^+\ell^-$ decay. From Ref.~\protect\cite{AtlasPhysTDR}.
\label{g1a_mllg}}
\end{figure}

If the NLSP is the $\lsp$ and it decays promptly, $\lsp \to \tG\gamma$, then
SUSY events contain two hard, isolated photons in addition to $\Etmiss$, jets,
and perhaps leptons. The decay chain $\neu2 \to \tell^\pm\ell^\mp \to
\lsp\ell^+\ell^- \to \tG \ell^+\ell^-\gamma$ provides, in addition to an
$\ell^+\ell^-$ endpoint like Fig.~\ref{c5_mll}, precisely measurable
$\ell\ell\gamma$ and $\ell\gamma$ endpoints. An example is shown in
Fig.~\ref{g1a_mllg}. These measurements alone allow the masses involved to
be determined precisely~\cite{AtlasPhysTDR}.

\begin{figure}
\dofig{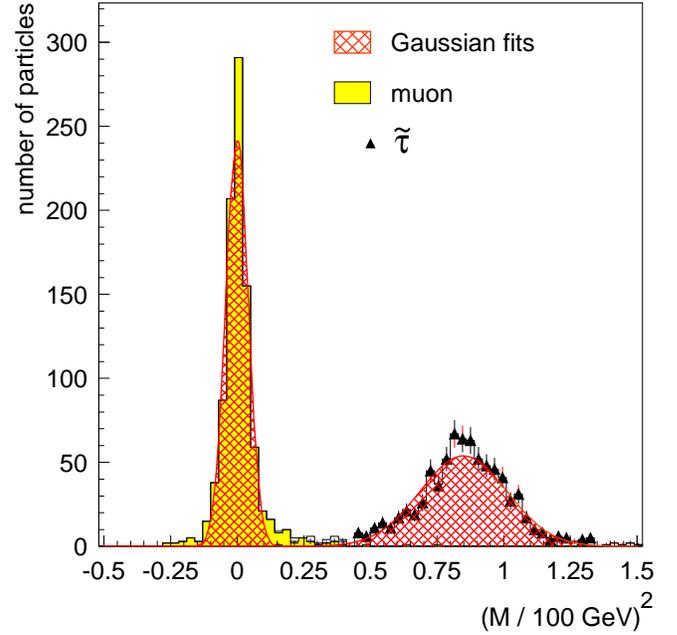}
\caption{Muon and $\s\tau_1$ masses reconstructed by time of flight.
From Ref.~\protect\cite{cmsnlsp}. \label{cmsstau}}
\end{figure}

If the NLSP is a $\ttau$ and is long-lived, then it penetrates the calorimeter
like a high momentum muon but has $\beta<1$. The $\ttau$ mass can be measured
directly using the muon chambers as a time-of-flight
system~\cite{AtlasPhysTDR, cmsnlsp}; see Fig.~\ref{cmsstau}. Once this mass
is known, all the other masses can be determined directly by observing mass
peaks~\cite{AtlasPhysTDR}.

The lifetime of the NLSP measures the overall SUSY breaking scale
and so is a crucial parameter in GMSB models. For a $\lsp$ NLSP with a very
short lifetime, the Dalitz decay $\lsp \to \tG e^+e^-$ can be used; the reach
is limited only by the resolution of the vertex detector. A long-lived $\lsp$
that decays inside the tracker will produce a photon that does not point to
the primary vertex. The ATLAS electromagnetic calorimeter provides pointing
and can detect such decays for $c\tau \simle 100\,{\rm
km}$\cite{AtlasPhysTDR}. The lifetime of a $\ttau$ can be
measured for $1 \simle c\tau \simle 100\,{\rm m}$ by counting the numbers of
events with one and two reconstructed sleptons~\cite{Ambrosanio:2000zu}. It
should also be possible to reconstruct $\ttau \to \tG \tau$ decays in the
central tracker.

\subsection{AMSB Measurements}

In the AMSB model the $\tchi_1^\pm$ and (mainly wino) $\lsp$ are almost
degenerate, while the (mainly bino) $\neu2$ is heavier. Hence, signatures
involving $\neu2$ decays are largely unchanged from similar SUGRA
ones~\cite{Paige:1999ui}. Typically, the splitting between the $\tchi_1^\pm$
and $\lsp$ is a few hundred MeV, so the chargino decays via $\tchi_1^\pm \to
\lsp \pi^\pm$ with $c\tau \sim 1\,{\rm cm}$ and is mostly invisible. The
fraction of single lepton events is consequently reduced. A small fraction of
the $\tchi_1^\pm$ will travel far enough to be seen in the vertex detectors.

\begin{figure}[t]
\dofig{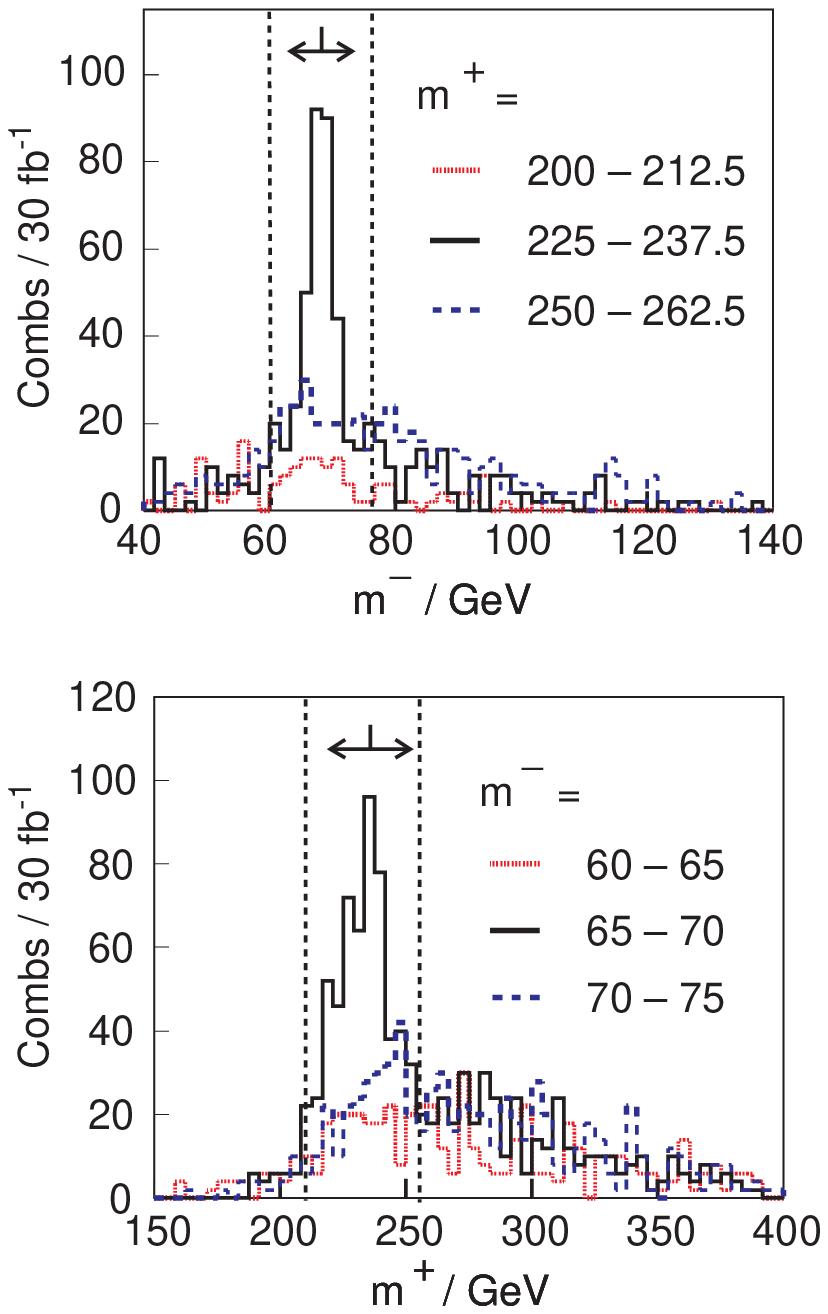}
\caption{Distributions of $m^\pm = M(\ell^+\ell^-qqq) \pm M(qqq)$ in
$R$-parity violating SUSY events with $\lsp \to qqq$ after selecting either
the peak or sidebands in $m^\mp$. From Ref.~\protect\cite{Allanach:2001xz}.
\label{camrviol}}
\end{figure}

\subsection{$R$-Parity Violation}

The SUGRA, GMSB, and AMSB models assume that $R$ parity is conserved so that 
the LSP is stable. It is possible that either baryon number or lepton number is
violated, allowing the LSP to decay; violation of both would allow rapid
proton decay. If lepton number is violated, then SUSY events will contain
multiple leptons, e.g., from $\lsp \to \ell^+\ell^-\nu$ or $\lsp \to \ell q
\bar q$. These cases are easy to detect, and similar partial reconstruction
techniques can be used~\cite{AtlasPhysTDR}.

If baryon number is violated, the LSP will decay into jets, $\lsp \to qqq$,
giving events with very high jet multiplicity and no (large) $\Etmiss$. The
QCD background for this is not well known, but it appears difficult to extract
the signal using only jets. It is possible, however, to reconstruct SUSY
events using cascade decays involving leptons. The results for an analysis
at a point with the decay chain $\neu2 \to \tell^\pm\ell^\mp \to \lsp
\ell^+\ell^- \to qqq\ell^+\ell^-$ is shown in Fig.~\ref{camrviol}. The mass
combinations $m^\pm = M(qqq\ell^+\ell^-) \pm M(qqq)$ show clear peaks
corresponding to the $\lsp$ and $\tchi_2^0$ masses. The $\tchi_2^0$,
$\tell_R$, and $\lsp$ masses can be determined from these plus a dilepton edge
similar to Fig.~\ref{c5_mll}.

\begin{figure}[t]
\dofig{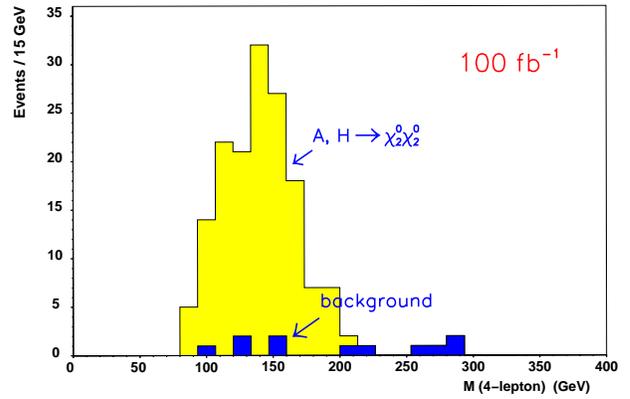}
\caption{
The invariant mass distribution of $\ell^+\ell^-\ell^+\ell^-$
  for leptons arising from the decay  $A\to \tchi_2^0\tchi_2^0$ for
  the MSSM with  $M_2=120$ GeV, $M_1=60$ GeV, $\mu =500$ GeV and
  $m_{\tilde{l}}=250$ GeV.
From Ref.~\protect\cite{moortgatref}.
\label{moortgat}}
\end{figure}

\begin{figure}
\dofig{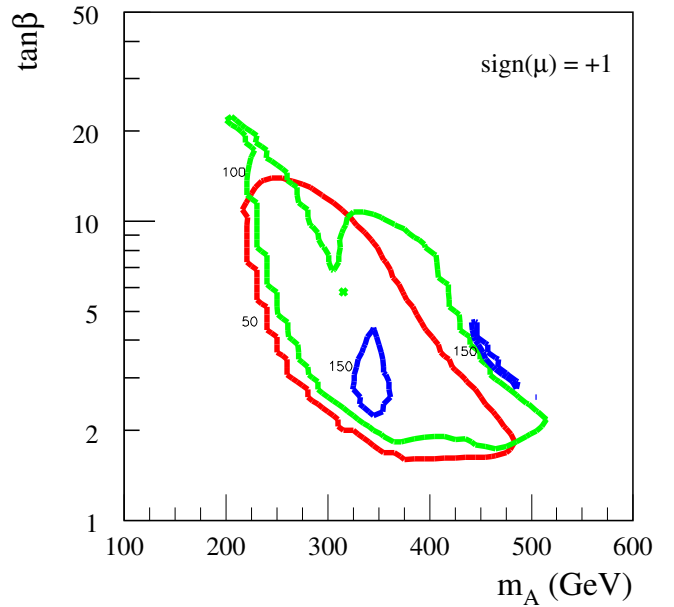}
\caption{The region in $M_A-\tan\beta$ where the decay $A\to
\tchi_2^0\tchi_2^0\to 4\ell+X$ is observable in the SUGRA model. The
contours  are labeled by $m_0$. An integrated luminosity of $300 \,
\fbi$ is assumed.
From Ref.~\protect\cite{AtlasPhysTDR}.
\label{achichi}}
\end{figure}

\subsection{Decay of Higgs to Sparticles}

For certain choices of the MSSM parameters, it is possible for the
heavy Higgs bosons $A$ and $H$ to decay to sparticles. As an example,
The decay $A,H\to \tchi_2^0\tchi_2^0$ has  been investigated by both
collaborations. The subsequent
decay $\tchi_2^0\to\ell^+\ell^-\lsp$ gives rise to events with four
isolated leptons. The invariant mass of the 4-lepton system for one
such case is shown in Fig.~\ref{moortgat}. Here the study
\cite{moortgatref} is done in the context of the MSSM. 

This signal is visible over a large fraction of parameter space as can
be seen from Fig.~\ref{achichi} which shows the accessible region in
the SUGRA model for various values of $m_0$. Note that the
value of $m_{1/2}$ is determined once $M_A$ and $m_0$ are given.
For large values of $M_A$, the decay $A,H\to
t\overline{t}$ dominates and the signal is unobservable. 

\subsection{SUSY Summary}

If SUSY with $R$ parity conservation exists at the TeV scale, then observation
of $\Etmiss$ plus multijet signatures with the ATLAS and CMS detectors at the
LHC should be straightforward. Many GMSB models provide additional handles.
If lepton number is violated, the signatures are easier. If baryon number is
violated, discovery probably must rely on selecting particular cascade decays,
although measurements are then easier. The kinematics and qualitative features
of the discovery signatures can be used to establish the approximate mass
scale and to distinguish classes of models.

If $R$ parity is conserved, then all SUSY decays contain a missing LSP, so no
mass peaks can be reconstructed. Kinematic endpoints of mass distributions
have proved useful for a number of SUSY points in a variety of SUSY
models~\cite{AtlasPhysTDR}. The method seems fairly general:  there is usually
at least one distinctive mode --- typically $\neu2 \to \neu1 \ell^+\ell^-$,
$\neu2 \to \tell_R^\pm\ell^\mp$, or $\neu2 \to \neu1 h \to \neu1 b \bar b$ ---
from which to start. These can be combined with jets to determine other
combinations of masses.

The SUSY events will contain much more information than just endpoints like
those described above. For example, while it is not possible to reconstruct
$\s\chi_{1}^\pm$ decays in the same way because of the missing neutrino, one
can get information about the chargino mass by studying $M_{\ell q}$ and other
distributions for 1-lepton events. Cross sections and branching ratios can
also be measured; interpretation of these will be limited by the theoretical
errors on the calculation of cross sections and acceptances. Without real
experimental data, it is difficult to assess such theoretical systematic
errors.

This program will provide a large amount of information about gluinos,
squarks, and their main decay products, including $\neu1$, $\neu2$,
$\tchi^\pm$, and any sleptons that occur in their decays. The heavy gauginos
typically have small cross sections, as do sleptons produced only by the
Drell-Yan process. High precision measurements of the LSP mass and of
couplings and branching ratios also appear more difficult.

\section{Strong EWSB Dynamics}

While the existing precision electroweak measurements are consistent with a
light Higgs boson, the possibility of electroweak symmetry breaking by new
strong dynamics at the TeV scale cannot be excluded.

\subsection{Strongly interacting $W$'s}

The couplings of longitudinally polarized gauge bosons to each other 
 are fixed at low energy
by the nature of the spontaneously broken electroweak symmetry 
and are independent of the details of the
breaking mechanism. Scattering amplitudes calculated from these 
couplings will violate 
unitarity at center of mass energies
of the $WW$ system  around 1.5 TeV. New physics must enter to cure 
this problem.
In the minimal Standard Model and its supersymmetric version, the cure
arises from the weakly coupled
Higgs bosons.   
If no Higgs-like particle exists, then new  non-perturbative dynamics
 must enter in the 
scattering amplitudes for $WW$, $WZ$ and $ZZ$ scattering at high energy.
Therefore, if no new physics shows up
at lower mass scales, one must be able to probe $W_L W_L$ scattering at 
$\sqrt{\hat s} \sim 1\,\TeV$. 

Various models exist that can be used as benchmarks for this
physics\cite{chanowitz}. The basic signal for all of these models is an
excess of events over that predicted by the Standard Model for gauge
boson pairs of large invariant mass. In certain models resonant
structure can be seen; an example of this is given in the next
subsection.  Since in the Standard Model there is no process
$q\overline{q}\to W^\pm W^\pm$, the $W^\pm W^\pm$ final state expected
to have a much smaller background than the $ZZ$ or $W^+W^-$ ones.
There are small $W^\pm W^\pm$ backgrounds from higher order processes
and from $WZ$ if one lepton is lost. The background from charge
misidentification is negligible in either ATLAS or CMS.

\begin{figure}[t]
\dofig{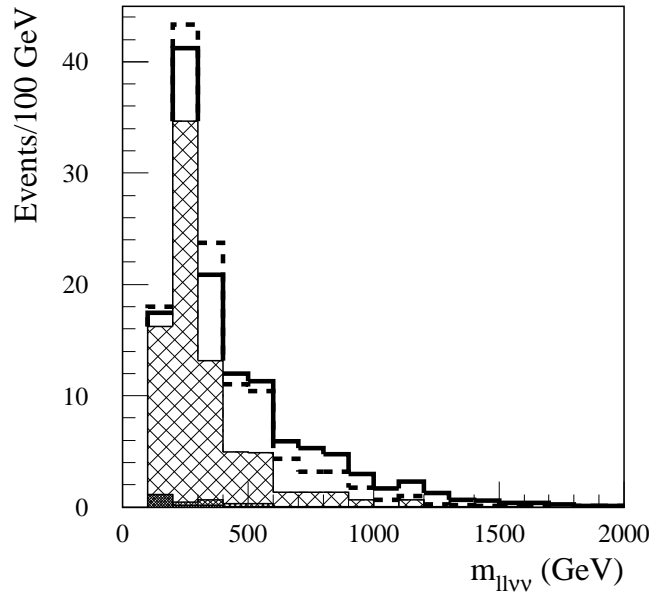}
\caption[]{The invariant mass  spectrum for same sign dileptons in 
the search for a 
strongly coupled $WW$ sector as simulated by ATLAS.
The signal corresponds to a $1\,\TeV$ Higgs boson.  The backgrounds
are from  $WZ$ and $W_TW_T$ production via electroweak bremsstrahlung.
From Ref.~\cite{AtlasPhysTDR}.\label{atlasww}}
\end{figure}

ATLAS \cite{AtlasPhysTDR} conducted a study of the 
signal and background in this channel. Events were selected that have two leptons of
the same sign with $p_T>40$ GeV and $\abseta <1.75$. If a third 
lepton was present that, in combination with one of the other two,
was consistent with the decay of a $Z$ (mass within 15 GeV of the $Z$ 
mass), the event was rejected. This cut is needed to
eliminate the background from $WZ$ and $ZZ$ final states. 
In addition the two leptons
are required to have invariant mass above 100 GeV and 
to be separated in $\phi$ so that $\cos\phi <-0.5$. At this stage, 
there are $\sim 1700$ Standard Model events for a luminosity of 
$300\,\fbi$. Of these events roughly 90\% are from
$WZ$ and $ZZ$ final states and 10\% from $Wt\overline{t}$.
There are of order 300 signal events depending upon the model used for 
the strongly coupled gauge boson
sector.  Additional cuts are needed to reduce the background. A 
jet veto requiring no jets with $p_T> 50$ GeV and $\abseta <2$
is effective against the $Wt\overline{t}$ final state. The requirement 
of two forward jet tags each with $15<p_T<150$ GeV
and $\abseta >2$ reduces the $WW$, $ZZ$ and $WZ$ background.

The remaining background of $\sim 80$ events is dominated by the
$q\overline{q}\to WW q\overline{q}$ processes. The signal rates vary
between 35 and 9 events depending upon the model. The largest rate
arises from a model where the $WW$ scattering amplitude, which is known
at small values of $\sqrt{s}$ from low energy theorems, is extrapolated
until it saturates unitarity and its growth is then cut off. 
The case of a 1 TeV Standard Model Higgs boson is shown in
Fig.~\ref{atlasww} where there are approximately  20 events.
It can be seen that the signal and background have
the same shape; therefore the establishment of a signal requires
confidence in the expected level of the background. The experiment is
very difficult, but at full luminosity, a signal might be extracted by
comparing the rate for $W^+W^+$ with those for $WZ$, $W^+W^-$, and $ZZ$
final states.

A similar study in CMS of the $W^+W^+$ final state leads to a similar
conclusion~\cite{smith96}.
Jet tagging (vetoing) in the forward (central)
region is essential to extract a signal.

\subsection{Technicolor}

Many models of strong electroweak symmetry breaking
(technicolor~\cite{Dimopoulos:1979es}\cite{Weinberg:1979bn},
topcolor-assisted technicolor~\cite{Hill:1991at}, 
BESS~\cite{bess}) predict resonances which decay 
into vector bosons (or their longitudinal components).  These signals are very
striking since they are produced with large cross sections and may be observed
in the leptonic decay modes of the $W$ and $Z$ where the backgrounds are very
small.  

\begin{figure}[t]
\dofig{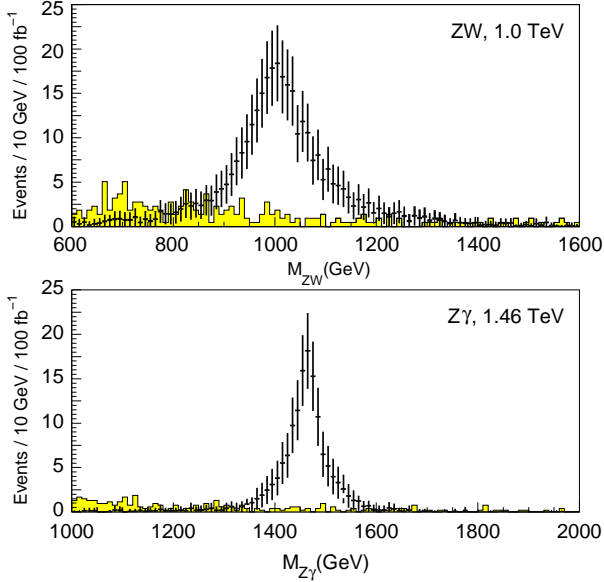}
\caption[]{Reconstructed masses for high-mass resonances decaying
into gauge boson pairs a simulated by ATLAS: $(a)$ $\rho_T$ of mass 1.0~TeV 
decaying into $WZ$ and subsequently into 3 leptons; and 
$(b)$ $\omega_T$ of mass 1.46~TeV decaying into
$Z\gamma$ with $Z \to 2$~leptons. From Ref.~\cite{atlas}. 
\label{atlastechni}}
\end{figure}

ATLAS has studied a techni-rho, $\rho_T \to WZ$, with
$W\to\ell\nu$, $Z\to\ell\ell$, for $m_{\rho_T} = 1.0$~TeV and also a
techni-omega, $\omega_T \to Z\gamma$, with
$Z\to\ell\ell$, for $m_{\omega_T} = 1.46$~TeV.  The backgrounds due to
$t\overline t$ and continuum vector-boson pair production are small as can be
seen in Fig.~\ref{atlastechni}.  

More challenging are the possible decays into non-leptonic modes such as
$\rho_T \to W (\ell\nu) \pi_T (b\overline b)$, which has a signature like
associated $WH$ production with $H \to b\overline b$;
$\eta_T \to t \overline t$, for which the signature is a resonance in
the $t \overline t$ invariant mass; and $\rho_{T8} \to$~jet~jet, 
for which the signature is a resonance in
the dijet invariant mass distribution.

\subsection{Compositeness}

\begin{figure}[t]
\dofig{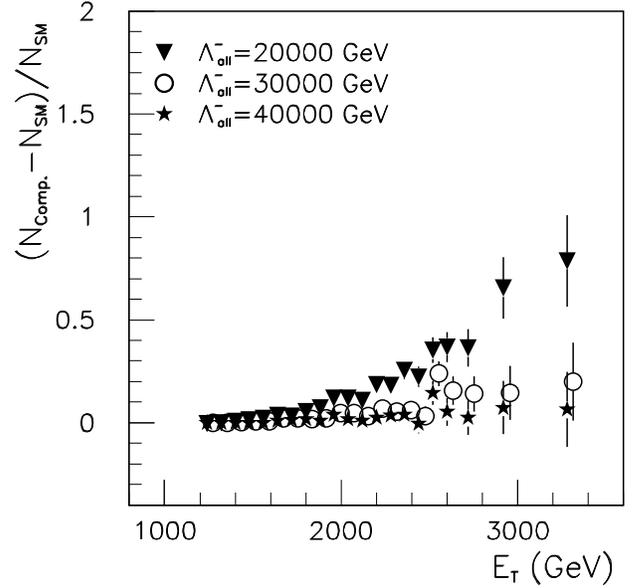}
\vskip-15pt
\caption[]{Difference of the Standard Model prediction and the effect of
compositeness on the jet $E_T$ distribution, normalized to the Standard
Model rate.  The errors correspond to 300 $\fbi$ for various values of
the compositeness scale $\Lambda$. From Ref.~\cite{AtlasPhysTDR}.
\label{atlas_compos}}
\end{figure}

If quarks have substructure, it will be revealed in the deviations
of the jet cross-section from that predicted by QCD. 
The deviation is parameterized by an interaction of the form
$$
{4\pi\over\Lambda^2} q\gamma^\mu\overline{q}\,q\gamma^\mu\overline{q}
$$
which is strong at a scale $\Lambda$. 
This is regarded as an effective interaction which is valid only 
for energies less than $\Lambda$.
The ATLAS collaboration has investigated the possibilities for searching for 
structure in the jet cross-section at high $p_T$.
Fig.~\ref{atlas_compos} shows the normalized jet 
cross section $d\sigma/dp_Td\eta$ at $\eta=0$. 
The rate is shown as a function of 
$p_T$ for various values of $\Lambda$ and is normalized to 
the value expected from QCD. The error bars at a particular  value of $p_T$
indicate the size of the statistical error to be expected at that value for 
luminosities of  $300$ $\fbi$.
It can be seen  that the LHC at full luminosity will be able to probe
up to 
$\Lambda=20$ TeV if the systematic uncertainties are smaller
than the statistical ones. Systematic effects are of two types;
theoretical 
uncertainties in calculating the QCD rates and detector
effects. The former are dependent upon an accurate knowledge of the
structure 
functions in the $x$ range of interest and upon
higher order QCD corrections to the jet cross-sections.  Uncertainties from 
these sources can be expected to be less than 10\% .

\begin{figure}[t]
\dofig{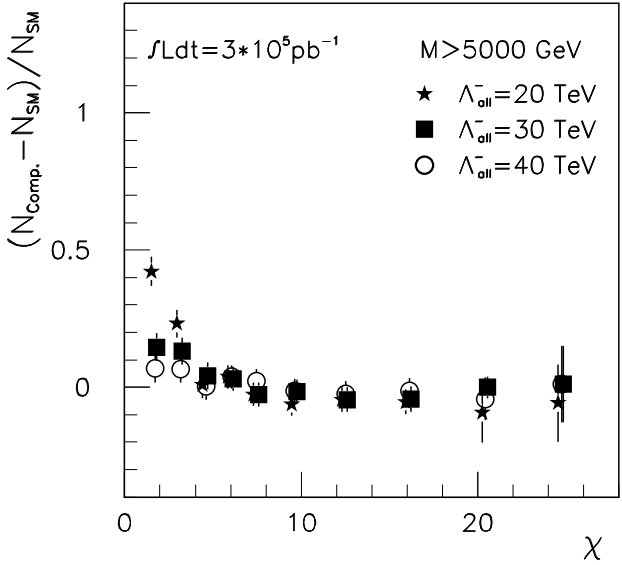}
\vskip-25pt
\caption[]{ Difference of the Standard Model prediction and the effect
of compositeness on the di-jet angular distribution for di-jet mass
above 5000 GeV, normalized to the Standard Model rate.  The errors
correspond to $300\,\fbi$ for various values of the compositeness scale
$\Lambda$. From Ref.~\protect\cite{AtlasPhysTDR}. 
\label{atlas_composangle}}
\end{figure}

Experimental uncertainties are of two types: mismeasurement due to
resolution 
and nonlinearities in the detector response. The former are
at the 20\% level; the latter can be more serious and can induce
changes in 
the apparent shape of the jet cross-section. In the case of ATLAS
these non-linearities could fake a compositeness effect with a scale
$\Lambda\sim 30\,\TeV$, which is beyond the limit of sensitivity.

The angular distribution of the jets in a dijet event selected 
so that the dijet pair has a very large mass is less sensitive 
 to the non-linearities. Events are selected with the invariant mass of
 the jet pair is above some $M_0$, and the variable $\chi$ defined by 
$\chi=(1+\cos\theta)/(1-\cos\theta)$ 
where $\theta$ is the angle of
an outgoing jet relative to the beam direction in the center of mass
frame of the jet pair. The distribution shown in
Fig.~\ref{atlas_composangle} illustrates that $\Lambda\sim 40 $ TeV is
accessible via this variable.

A better constraint on the scale $\Lambda$ may be obtained from
Drell-Yan dilepton final states, if leptons and quarks are both
composite and share common constituents.

\section{New Gauge Bosons}

A generic prediction of superstring theories is the 
existence of additional $U(1)$ gauge groups.  There is thus
motivation to search for additional $W^\prime$ and
$Z^\prime$ bosons.  
The current Tevatron
limit is 720~GeV for $W^\prime$ \cite{tevzprime}.

ATLAS and CMS have studied the sensitivity to a new neutral $Z^\prime$ boson
in $e^+e^-$, $\mu\mu$ and jet-jet final states, for
various masses and couplings \cite{atlaszprime,cmszprime}.
It is assumed that $\Gamma_{Z^\prime}\propto m_{Z^\prime}$.
ATLAS finds the best sensitivity in the $e^+e^-$ mode, in which 
signals could be seen up to $m_{Z^\prime}=5$~TeV for Standard-Model couplings. 
The other final states would provide important information on the $Z^\prime$
couplings. 
The pseudorapidity coverage over which lepton identification and measurement
can be carried out is important for $Z^\prime$ searches: should a signal
be observed, the forward-backward asymmetry of the charged leptons would
provide important information on its nature.  ATLAS found that reducing the
lepton coverage from $|\eta|\leq 2.5$ to $|\eta|\leq 1.2$ roughly halved the
observed asymmetries and prevented discrimination between two particular
$Z^\prime$ models which they investigated.

\begin{figure}[t]
\dofig{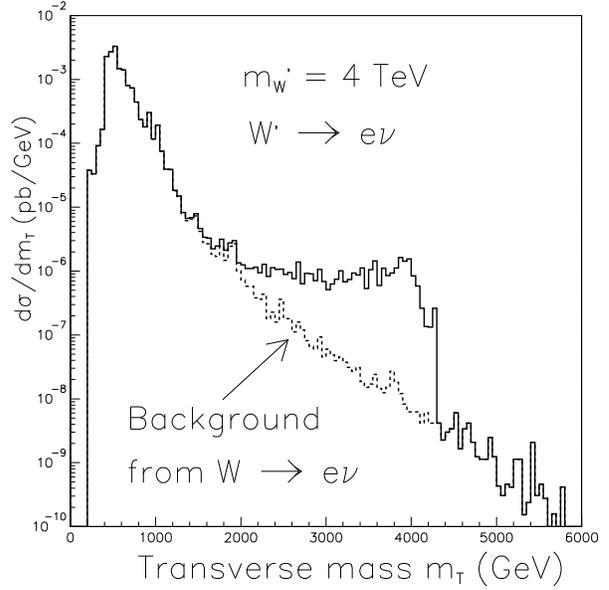}
\caption[]{Expected electron-neutrino transverse mass distribution in
ATLAS for $W^\prime \to e \nu$ decays with $M_{W^\prime}=4\,\TeV$ (solid)
above the dominant background from $W \to e \nu$ decays (dashed).
From Ref.~\cite{atlas}. \label{atlaswprime}}
\end{figure}

ATLAS also investigated their sensitivity to a new charged boson
$W^\prime$ decaying into $e\nu$.  The signal is structure in the transverse
mass distribution at masses much greater than $m_W$.  Fig.~\ref{atlaswprime}
shows the signal for a 4~TeV $W^\prime$.  They conclude that with
$100 \fbi$ one would be sensitive to $m_{W^\prime}=6$~TeV
and that the mass could be measured with a precision of  50 GeV.
Similar results for the sensitivity to new $W'$ bosons have also
been obtained by CMS \cite{cmswprime}.

\section{Extra Dimensions}

\begin{figure}[t]
\dofig{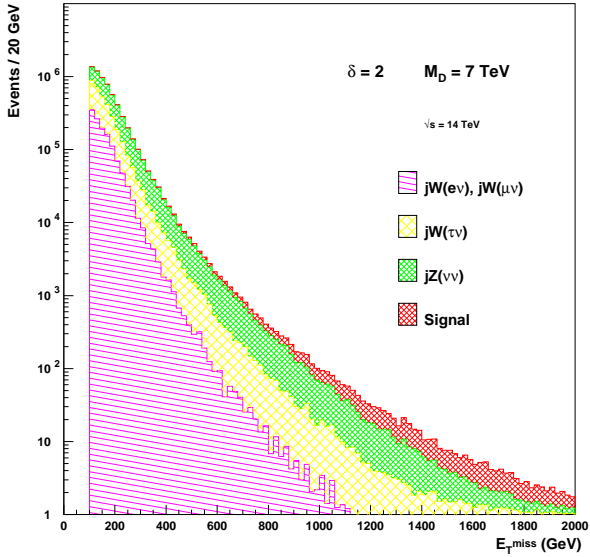}
\caption[]{Distributions of the missing transverse energy 
  in extra dimensions signal and in background events 
  after the selection and for 100 fb$^{-1}$ of integrated luminosity.  
   $\delta =2,M_{D}=7$ TeV is shown for the signal.
From Ref.~\cite{Vacavant:2000wz}.\label{extrad1}}
\end{figure}

There is much recent theoretical interest in models of
particle physics that have extra-dimensions in addition to the 3+1
dimensions of normal
space-time~\cite{Arkani-Hamed:1998,otherextrad,Randall:1999ee}.
In these models, new
physics can appear at a mass scale of order 1 TeV and can therefore be 
accessible at LHC. Two generic types of signals have been discussed.
In models of large extra-dimensions~\cite{Arkani-Hamed:1998}, there is
a tower of states consisting of massive graviton excitations whose
properties are 
parameterized in terms of two parameters, the number 
$\delta$ of additional
dimensions and  the fundamental scale $M_D$. The size of the extra
dimensions $R$ can be expressed in terms of these. 
Graviton excitations are produced in quark or gluon scattering; since
they have gravitational strength couplings, they escape the detector,
giving rise to final states with jets or photons plus missing transverse
energy.
Backgrounds
arise from the production of $Z$ or $W$ in association with a 
jet~\cite{Vacavant:2000wz}.
Fig.~\ref{extrad1} shows the distribution in missing  transverse energy
for the signal and background. The signal is manifest as an excess at
large $\Etmiss$. For $\delta=2$, and an  integrated luminosity of 100
$\fbi$ values of $M_D$ less than 9 TeV are accessible. The signal
could be confirmed from the $\gamma+\Etmiss$ channel as the rates in
this channel are predicted in terms of the same parameters.

\begin{figure}[t]
\dofig{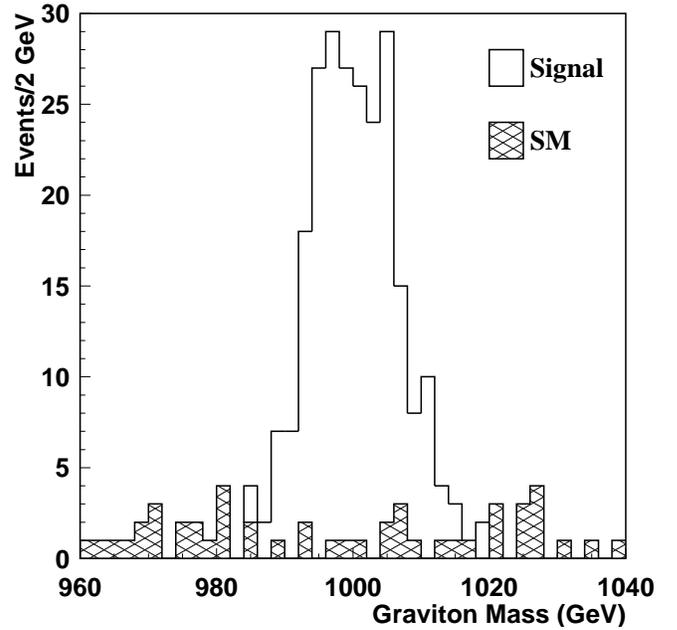}
\caption[]{Distributions of the $e^+e^-$ invariant mass
in signal from a graviton resonance of mass 1 TeV and in background 
after event selection and for 100 fb$^{-1}$ of integrated luminosity.  
From Ref.~\cite{Allanach:2000nr}. \label{extrad2}}
\end{figure}

In models of small (warped) extra dimensions~\cite{Randall:1999ee}, the
graviton excitations are much more massive and decay into jets,
leptons or photons. The decay into leptonic final states has been
studied~\cite{Allanach:2000nr}. Signals are  similar to those of
new gauge bosons except that that graviton resonances have
spin-2. Fig.~\ref{extrad2} shows how such a resonance would appear in
the $e^+e^-$ mass distribution. The signal  is visible for gravitons
where $\sigma\times B \simge 0.5\,\fb$ or approximately 2 TeV in the
model used in Ref.~\cite{Allanach:2000nr}. Confirmation that the signal
is indeed a graviton comes from measurements of the angular
distribution that confirms that the resonance is spin-2 and possible
observation in other final states such as $\gamma\gamma$.

\section{Anomalous Gauge-Boson Couplings}

The trilinear $WWV$ and $Z\gamma V$ couplings ($V=Z, \gamma$) may be probed at
hadron colliders using diboson final states.  
Following the usual notation,
the CP-conserving $WWV$ anomalous couplings are parameterized 
by five parameters:
$\Delta\kappa_Z$, $\lambda_Z$, $\Delta\kappa_\gamma$, 
$\lambda_\gamma$ and $\Delta g_1^Z$~\cite{peccei}. In the Standard Model,
$\kappa_{Z,\gamma}=1$, $\Delta g_1^Z=1$ and $\lambda_{Z,\gamma}=0$
In general, we would expect anomalous couplings of order $m_W^2/\Lambda^2$ if
$\Lambda$ is the scale for new physics, so if $\Lambda\sim 1\,$TeV then
$\Delta\kappa_V,\lambda_V\sim 0.01$.

To maintain
unitarity,  anomalous couplings must be modified by a form 
factor; so (for example) 
\begin{equation}
\Delta\kappa_V(q^2) = {\Delta\kappa_V^0 \over (1 + q^2/\Lambda_{FF}^2)^n}
\end{equation}
where $\Lambda_{FF}$ is the form factor scale and $n=2$ for $\Delta\kappa,
\lambda$.

The ATLAS collaboration has studied~\cite{upgrade} the sensitivity 
to anomalous couplings in the $W\gamma$ and $WZ$ modes;
the $W^+W^-$ signal is swamped by $t\overline t$ background. A 
form factor scale $\Lambda_{FF}=10\,$TeV was used.
For the $W\gamma$ final state, events were assumed to be triggered
using a high-$p_T$ lepton plus a high-$p_T$ photon candidate.  The background
includes contributions from events with a real lepton and a real photon (e.g.
$b\overline b \gamma$, $t \overline t \gamma$, and $Z\gamma$); a fake lepton 
but a real photon (e.g. $\gamma+{\rm jet}$); and a fake photon with a real
lepton (e.g. $W+{\rm jet}$, $b\overline b$, and $t\overline t$).  
Rejection factors of $10^4$ against jets faking photons and 
$10^5$ against jets faking electrons  were used (consistent with the
results from full simulation).  
To reduce backgrounds, events
were selected with 
$p_T^\gamma > 100\,{\rm GeV}$, $p_T^\ell > 40\,{\rm GeV}$,
and $|\eta^\ell|<2.5$.  Events with jets were also vetoed, to further reduce 
backgrounds and to lessen the importance of higher-order QCD corrections.
In an integrated luminosity of $100 \fbi$, 7500 events remain, with a
signal to background ratio of 3:1.  The $p_T^\gamma$ distribution is
then
 fitted
in the region where the Standard Model prediction is 15 events (above about 
$600\,{\rm GeV}$), yielding limits of $|\Delta\kappa_\gamma| < 0.04$ and
$|\lambda_\gamma| < 0.0025$ (95\% C.L.).

Similar techniques were used for the $WZ$ state.  The trigger was three
high-$p_T$ leptons, and the backgrounds are from $Zb\overline b$, $Z+{\rm jet}$,
$b \overline b$ and $t\overline t$ processes.  Events were selected with
$p_T^\ell > 25\,{\rm GeV}$, $|\eta^\ell|<2.5$, 
$|m_{\ell_1 \ell_2} - m_Z|< 10\,
{\rm GeV}^2$, and $m_T(\ell^3, \Etmiss) > 40\,{\rm GeV}^2$; a jet veto
was also imposed.  In $100 \fbi$, 4000
events then remain, with a
signal to background ratio of 2:1.  The $p_T^Z$ distribution is again fitted
in the region where the Standard Model prediction is 15 events (above about
$380\,{\rm GeV}$), yielding limits of $|\Delta\kappa_Z| < 0.07$ and
$|\lambda_Z| < 0.005$ (95\% C.L.).

\begin{figure}
\dofig{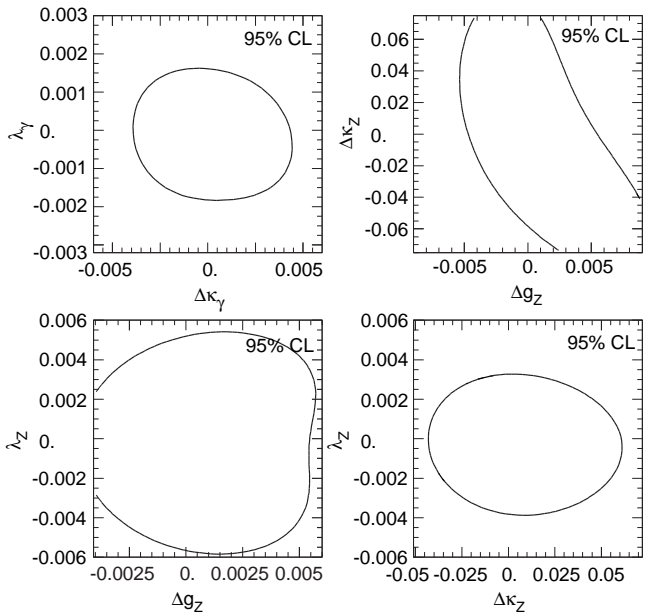}
\caption[]{95\% CL sensitivity limits on anomalous couplings
from $W\gamma$ and $Z\gamma$ production for an integrated luminosity
of 100 $\fbi$. From Ref.~\cite{upgrade}.\label{ellipseone}}
\end{figure}

 A  likelihood fit to the  distributions then yields correlated limits on
$\Delta\kappa_Z$, $\lambda_Z$,  $\Delta g_1^Z$, $\Delta\kappa_\gamma$ and
$\lambda_\gamma$ 
 which are shown in Fig~\ref{ellipseone}. These limits are comparable
 to deviations expected from radiative corrections in the Standard
 Model and extensions thereof~\cite{kappa-rad}. Better precision might
 be obtained by using the angular distributions.

\section{Standard Model Physics}

\subsection{Top Quark Physics}

The potential for the study of the top quark at hadron colliders 
is already apparent. 
The LHC will be a top factory, with 
about 10$^7$ $t\overline t$ pairs produced per year at a luminosity
of $10^{33}\,{\rm cm}^{-2}{\rm s}^{-1}$.  This would result in about
200,000 reconstructed $t\overline t \to (\ell\nu b)(jjb)$ events and
20,000 clean $e\mu$ events.  

\subsubsection{Top Mass Measurement}

The top mass can  be reconstructed from the $t\overline t \to (\ell\nu b)(jjb)$
final state using the invariant mass of the 3-jet system.  Problems arise
from  systematic
effects due to the detector and the theoretical modeling of the
production dynamics.  This measurement requires, of course, that the hadronic 
calorimetry be
calibrated to this level in the absolute energy scale and that its response be
stable over time.
ATLAS~\cite{AtlasPhysTDR} has studied these effects and concludes  that an 
accuracy of better than $\pm 2$~GeV could be attained. A complementary
method exploits very high-$p_T$ top quarks, where the decay products are boosted
and thus close.  Combinatorics and uncertainties associated with
measuring the individual jets are reduced, whereas those from jet
energy calibrations are increased with the result that the expected
errors are comparable.

The mass may also be reconstructed from dilepton events.  ATLAS estimates that,
by selecting events with two leptons from $W$ decays and an additional lepton 
from $b$-decay, and plotting the invariant mass of the lepton pair originating
from the same top decay, the mass could be determined with a statistical
accuracy of $\pm 0.5$~GeV, and a total accuracy of about $\pm 2$~GeV. The
dominant systematic errors arise from uncertainties in the $b$-quark
fragmentation and are therefore complementary to the 3-jet system
which is dominated by calorimeter and jet systematics.

\subsubsection{Rare Top Decays}

The large statistics available at LHC will provide sensitivity to other
non-standard or rare top decays.  As an example, ATLAS have investigated
the channel $t \to Zc$~\cite{AtlasPhysTDR}, which should occur at 
a negligible level in the SM. 
With an integrated luminosity of $100\,\fbi$, branching ratios
as small as $5\times 10^{-5}$ could be measured.

It has been estimated that ~\cite{tev2k} LHC will 
attain a precision 2--3 times
better than that ultimately achievable at the Tevatron on the ratio of longitudinal to left-handed $W$'s produced in
$t$ decays.  This ratio is exactly predicted in the SM for a given top mass,
and is sensitive to non-standard couplings at the $t \to Wb$ vertex, such
as a possible $V+A$ contribution.  

\subsection{$B$ Physics}

The preceding sections have shown the importance of $b$-tagging in addressing
many of the high-$p_T$ physics goals of the LHC.  Both major detectors will
consequently have the capability to tag heavy flavor production through
displaced vertices.  This capability, together with the large
$b$-quark production cross-section at the LHC, 
will enable them to also pursue an interesting program of
$B$-physics.  It can be assumed that CP violation in the $b-$quark 
system will have been observed before the LHC gives data.
Nevertheless the enormous rate will enable a very precise
determination 
of $\sin 2\beta$ to be made using the decay $B_d\to
\Psi K_S$ ($\Psi \equiv J/\psi, \psi(2S)$). An error of $\pm 0.02$ can be expected after $10$ fb$^{-1}$
of integrated luminosity. It will also be possible to
measure $B_s\overline{B}_s$ mixing and to search for rare decays such 
as $B\to \mu\mu$.  

\section{Summary and Conclusions}

The $SU(3) \times SU(2) \times U(1)$ gauge interactions of the Standard Model
provide an elegant and a tremendously successful description of
existing data, but they give no explanation of the
origin of particle masses. The internal consistency of the Standard Model
requires that at least part of the explanation of masses, the origin of
electroweak symmetry breaking, must be found at the TeV scale. The LHC is
unique among accelerators currently existing or under construction in that it
has sufficient energy and luminosity to study that mass scale in detail.  More
specifically, the very detailed simulation studies carried out by the ATLAS
and CMS collaborations enable one to make the following statements with a high
degree of confidence:

\begin{itemize}
\item If the minimal Standard Model is correct and the Higgs boson is not
discovered previously, it will be found at LHC.

\item If supersymmetry is relevant to the breaking of electroweak symmetry, it
will be discovered at LHC and many details of the particular supersymmetric
model will be disentangled.

\item If the Higgs sector is that of the minimal supersymmetric model, at
least one Higgs decay channel will be seen, no matter what the parameters turn
out to be. In many cases, several Higgs bosons or decay channels will be
seen.

\item If the electroweak symmetry breaking proceeds via some new strong
interactions, many resonances and new exotic particles will almost certainly
be observed.

\item New gauge bosons with masses less than several TeV will be discovered or
ruled out.

\item Signals for extra-dimensions will be revealed if the relevant scale is in
the TeV range.
\end{itemize}

\noindent The LHC represents a great opportunity --- and promise of vast
excitement --- not only for the collaborators on the LHC experiments but for
the whole field of particle physics.

\bigskip

We thank our many ATLAS and CMS colleagues who have carried out the
work summarized here.

The work was supported in part by the Director, Office of Energy
Research, Office of High Energy Physics, Division of High Energy
Physics of the U.S. Department of Energy under Contracts
DE--AC03--76SF00098 and DE-AC02-98CH10886.  Accordingly, the U.S.
Government retains a nonexclusive, royalty-free license to publish or
reproduce the published form of this contribution, or allow others to
do so, for U.S. Government purposes. 


\end{document}